\documentclass[preprint,showpacs,preprintnumbers,amsmath,amssymb,pra,letter,floatfix]{revtex4}

\usepackage{graphicx}
\usepackage{dcolumn}
\usepackage{bm}

\begin{document}


\title{Scalar and vector modulation
instabilities induced by vacuum fluctuations in fibers:\\
numerical study}

\author{E. Brainis}
\email{Edouard.Brainis@ulb.ac.be}
\author{D. Amans}%
\email{David.Amans@ulb.ac.be}
\affiliation{%
Optique et acoustique, Universit\'e libre de Bruxelles,%
Avenue F.D. Roosvelt 50, CP 194/5, 1050 Bruxelles, Belgium%
}%
\author{S. Massar}%
\email{smassar@ulb.ac.be}
\affiliation{%
Service de physique th\'{e}orique, Universit\'{e} libre de %
Bruxelles, Boulevard du Triomphe, CP 225, 1050 Bruxelles, Belgium%
}%
\date{\today}

\begin{abstract}
We study scalar and vector modulation instabilities induced by the
vacuum fluctuations in birefringent optical fibers. To this end,
stochastic coupled nonlinear Schr\"{o}dinger equations are
derived. The stochastic model is equivalent to the quantum field
operators equations and allow for dispersion, nonlinearity, and
arbitrary level of birefringence. Numerical integration of the
stochastic equations is compared to analytical formulas in the
case of scalar modulation instability and non depleted pump
approximation. The effect of classical noise and its competition
with vacuum fluctuations for inducing modulation instability is
also addressed.
\end{abstract}

\pacs{42.50.Ct, 42.65.Sf}
\maketitle

\section{\label{sec:1}Introduction}
In the early 1980s single-mode silica optical fibers were
recognized as a privileged medium for experiments in quantum
optics because they exhibit a well defined transverse mode and
very low losses. The Kerr nonlinearity of silica has been used to
produce non classical states of light. Squeezed state production
in fibers was pioneered by Levenson \textit{et al.}
\cite{Levenson85a,Levenson85b} in 1985. Since then a large number
of squeezing experiments have been performed with cw-light or
short optical pulses (for a review see \cite{Sizmann99}).

More recently, Fiorentino \textit{et al.} demonstrated that
optical fibers may also be used to produce \emph{twin-photons}
pairs \cite{Fiorentino02}. This new kind of twin-photon source is
well suited for fiber-optic quantum communication and quantum
cryptography networks. In contrast with the traditional
twin-photon sources, based on the $\chi^{(2)}$ down-conversion
process, it avoids large coupling losses occurring when the pairs
are launched into long distance communication fibers. The physical
process used in \cite{Fiorentino02} to generate twin-photon pairs
is a four-wave mixing (FWM) phase-matched by the interplay of the
optical Kerr effect and the chromatic dispersion, called (scalar)
modulation instability (MI). This process can be basically
understood as the destruction of two pump photons at frequency
$\omega_0$ and the creation of Stokes (red-shifted) and
anti-Stokes (blue-shifted) photons at frequencies $\omega_s$ and
$\omega_a$ satisfying the energy conservation relation
$2\omega_0=\omega_s+\omega_a$. In the time domain, the beating of
the pump, signal and idler waves produces a fast modulation of the
pump envelope.

MI is a spontaneous phenomena which can be
described in the framework of classical nonlinear optics
\cite{Agrawal95}. In this classical framework, one usually considers
that MI is
induced by some incoherent noise initially present on the pump
wave. Twin-photon pair production requires however a very low
input noise power to be efficient (otherwise the twin-photons
are buried in the background photon noise). Such a regime is
dominated by vacuum fluctuations and requires a quantized field
theory to be described properly. A small-perturbation approach of
this problem \cite{Potasek87} (non depleted pump approximation)
shows that vacuum fluctuations can induce the MI even in the
absence of any classical input noise. This is an ideal situation
for twin-photon pairs production, but it never occurs in a real
life experiment. In practice, classical input noise and vacuum
fluctuations compete for inducing MI.

Understanding the dynamical development of MI from a quantum point
of view is an important issue for the design of fiber-optics
twin-photon pairs sources. We present an unified approach to this
problem based on the \emph{stochastic nonlinear Schr\"{o}dinger
equations} (SNLSE)
\cite{Carter87,Drummond87,Kennedy88a,Kennedy88b,Kennedy91,Drummond01a}.
This formalism is equivalent to quantum-field Heisenberg equations
but has two main advantages. First, it is very suitable for
numerical simulations of complex situations where classical noise
and vacuum fluctuations act together. Second the
\emph{correspondence principle} of quantum physics, i.e. the
transition from quantum to classical world, looks very natural in
the SNLSE formalism.

The MI observed in standard non-birefringent single-mode silica
fibers is often referred to as \emph{scalar} modulation
instability (S-MI) because polarization of light plays no role in
this process. In birefringent fibers different kinds of MI may
appear because of the interplay of nonlinearity, chromatic
dispersion and birefringence. This case will be referred as
\emph{vector} modulation instability (V-MI). Twin-photon pairs
sources based on V-MI have not been demonstrated yet, in contrast
with S-MI based sources \cite{Fiorentino02}. In this article, we
will however address both issues because we recognize that V-MI
twin-photon pairs sources could be more practicable than S-MI
based ones. The theoretical and computational tools that we
developed apply to all kind of MI, including the
low-birefringence, high-birefringence and scalar limits.

This article is divided into four sections, beginning with the
present introduction. The second section consists of a review of
the approach based on a perturbation analysis around the steady
state solution. This is the simplest method of approaching the
problem of MI. It allows us to set the stage for the more
sophisticated approach we then develop and to make contact with
earlier work in the field. Then in Sec.~\ref{sec:3} we exhibit the
SNLSE for scalar and vector MI. Contrary to the approach of
Sec.~\ref{sec:2} these equations are not based on a perturbation
analysis and are valid in the strongly non linear regime. They
describe both MI originating from classical noise and from vacuum
fluctuations (quantum noise). The SNLSE we obtain generalize the
earlier results of
\cite{Carter87,Drummond87,Kennedy88a,Kennedy88b,Kennedy91}. In
particular the SNLSE we obtain does not require the birefringence
to be either small or large as in earlier work
\cite{Kennedy88b,Kennedy91}, but are valid for all values of the
birefringence. For the interested reader we present a self
contained derivation of these equations in the Appendix. In
Sec.~\ref{sec:4} we use the \emph{split-step Fourier} method to
integrate the SNLSE derived in Sec.~\ref{sec:3}, and illustrate
our algorithm on the cases of scalar MI and of vector MI both for
weak, intermediate, and strong birefringence. In order to
interpret the results of the numerical integration it is essential
to introduce the notion of \emph{mode}. This allows us to compare
the numerical results and the analytical solutions derived from
perturbation theory. We also compare in detail the characteristics
of MI induced by classical noise and by quantum noise. In a
companion article \cite{Amans} we shall show that our numerical
results are in very good quantitative agreement with experimental
results.

\section{\label{sec:2}Scalar modulation instability: perturbation analysis}
As pointed out in the introduction several kind of MI can occur in
optical fibers. From a general point of view, MI occurs when the
continuous wave (steady-state) propagation is unstable. The most
straightforward way to get some insight on how the MI develops is
to perform a small-perturbation analysis of the steady-state.
However this method has limitations: it cannot address the
strongly non linear regime and simple analytical formulas cannot
be always obtained by this method.

In this section we illustrate the perturbation method in the case
of S-MI, which is the simplest. S-MI occurs in isotropic fibers in
the anomalous dispersion regime. In what follows we will review
the most important aspects of this approach, focussing on the
comparison between the classical and quantum descriptions, and on
the limitations of this approach. In Sec.~\ref{sec:4}, we will
compare the analytical formulas for S-MI derived from the
perturbation analysis to numerical results from the SNLSE derived
in Sec.~\ref{sec:3}. The discussion of MI in birefringent fibers
is also postponed to Sec~\ref{sec:3} and Sec.~\ref{sec:4}.

\subsection{\label{sec:2a}Classical description}
In an isotropic single-mode fiber, the electric field may be
written
\begin{equation}
\mathbf{E}(\mathbf{r},t)=F(x,y)A(z,t)\exp[i\beta_0z-\omega_0
t]\mathbf{\hat{x}}+\mathrm{c.c.}
\end{equation}
where $\mathbf{\hat{x}}$ is a unit vector orthogonal to the fiber
axis ($z$-axis), $\omega_0$ the carrier angular frequency and
$\beta_0=\beta(\omega_0)$ the associated propagation wave number
(modal propagation constant). $F(x,y)$ stands for the mode profile
function and $A(z,t)$ for the field envelope. The field is
supposed to be polarized linearly. This is not restrictive because
of the isotropy assumption. The complex envelope $A$ evolves
according to the \emph{nonlinear Shr\"{o}dinger equation} (NLSE)
that can be established from Maxwell's electromagnetic theory
\cite{Agrawal95}. If the envelope is normalized in such way that
$|A(z,t)|^2$ is equal to the instantaneous power flowing through
the plane $z=\text{constant}$ at time $t$, the NLSE is obtained in
the following form:
\begin{equation} \label{NLS}\frac{\partial A}{\partial
z}+\frac{1}{v_g}\frac{\partial A}{\partial
t}=-i\frac{\beta_2}{2}\frac{\partial^2A}{\partial t^2}+i\gamma
|A|^2A.
\end{equation}
Here $v_g=(d\beta/d\omega)^{-1}$ is the group velocity of the
wave, $\beta_2=d^2\beta/d\omega^2$ the group-velocity dispersion
(GVD) parameter and $\gamma$ is fiber nonlinearity parameter
defined as
\begin{equation}\label{gamma}
\gamma=\frac{3\omega_0\chi_{xxxx}}{4\epsilon_0
n_0^2c^2A_{\text{eff}}}
\end{equation}
where $n_0$ is the fiber mean linear index of refraction,
$A_{\text{eff}}$ the mode effective area and $\chi_{xxxx}$ is the
diagonal element of the fiber $\chi^{(3)}$ nonlinearity tensor.

S-MI is observed when a continuous wave (cw) or a quasi-cw optical
pulse is launched into the fiber. In the cw-case, when an optical
power $P_0$ at frequency $\omega_0$ in injected, a first order
perturbation analysis of Eq.~(\ref{NLS}) shows that the cw
steady-state solution $A_{st}(z)=\sqrt{P_0}\exp{(i\gamma P_0 z)}$
becomes instable in the anomalous dispersion regime ($\beta_2<0$).
The instability manifests itself by a parametric gain at
frequencies $\omega_0\pm\Omega$ with $0<\Omega<2\sqrt{\gamma
P_0/|\beta_2|}$. The maximal gain $g=2\gamma P_0 L$ occurs for
$\Omega=\sqrt{2\gamma P_0/|\beta_2|}\equiv\Omega_{\text{max}}$
\cite{Agrawal95}. Noise at these frequencies is strongly
amplified. As a result, the optical spectrum at the fiber output
exhibits two sidebands at frequencies
$\omega_0\pm\Omega_{\text{max}}$. This analysis supposes that the
pump power remains constant (undepleted pump approximation) and a
cw regime. However, numerical simulations of Eq.~(\ref{NLS}) show
that the above formula also hold for quasi-cw (for instance
nanosecond) optical pulses (see Sec.~\ref{sec:4}).

Sideband generation can be interpreted as a FWM process
phase-matched by the interplay of dispersion and Kerr
nonlinearity. From this point of view, the MI process can be seen
as the destruction of two pump photons at frequency $\omega_0$
followed by the creation of Stokes and anti-Stokes photons at
frequencies $\omega_s=\omega_0-\Omega_{\text{max}}$ and
$\omega_a=\omega_0+\Omega_{\text{max}}$ respectively. In the
cw-case and non depleted pump approximation, the output power
spectral density at frequencies $\omega_s$ and $\omega_a$ can be
computed using analytical formulas \cite{Stolen82} for given
initial conditions. These formulas also hold for quasi-cw pulses
provided one replace Stokes and anti-Stokes power spectral
densities by the number $n_s$ and $n_a$ of Stokes and anti-Stokes
photons located in the same temporal mode as the pump wave (see
Sec.~\ref{sec:4b} for a detailed analysis of this issue).
For example, assuming an incoherent initial
noise, one finds
\begin{subequations}\label{S-MIcl}
\begin{eqnarray}
n_s(L)&=&n_s(0)\cosh^2(\gamma P_0 L)+n_a(0)\sinh^2(\gamma P_0
L),\phantom{MM}\\
n_a(L)&=&n_a(0)\cosh^2(\gamma P_0 L)+n_s(0)\sinh^2(\gamma P_0
L),\phantom{MM}
\end{eqnarray}
\end{subequations}
where $L$ is the propagation distance. These equations show that
classical incoherent noise is amplified exponentially with gain
$g=2\gamma P_0 L$.

Noise induced S-MI was first demonstrated experimentally by Tai
\textit{et al.} in 1986 \cite{Tai86a}. This experiment confirmed
two main predictions of the classical theory: the amplification of
Stokes and anti-Stokes photons and the power-dependence of their
frequency shift. However the classical theory of S-MI only holds
when the initial number of noise photons is high or in a
stimulated regime when a coherent probe pulse at Stokes or
anti-Stokes frequency is injected together with the pump pulse
\cite{Tai86b}. Eqs.~(\ref{S-MIcl}) are unable to explain
\textsf{ab nihilo} generation of Stokes and anti-Stokes
twin-photons reported in \cite{Fiorentino02}. Neither are they
valid when the mean photon numbers $n_s(0)$ and $n_a(0)$ are of
the order of one or lower. In this regime vacuum fluctuations play
a central role and field quantization is required.

\subsection{\label{sec:2b}Quantum description}
To take into account vacuum fluctuations, fields must be
quantized. The quantum counterpart of the NLSE (\ref{NLS}) is
known as the quantum nonlinear Shr\"{o}dinger equation (QNLSE)
\cite{Kennedy88a,LaiHaus.I.89,Wright91,Haus00,Perina01}:
\begin{equation}\label{QNLS}
\frac{\partial \hat{A}}{\partial z}+\frac{1}{v_g}\frac{\partial
\hat{A}}{\partial
t}=-i\frac{\beta_2}{2}\frac{\partial^2\hat{A}}{\partial
t^2}+i\gamma\hat{A}^\dag\hat{A}\hat{A},
\end{equation}
where $\hat{A}$ is the quantum operator corresponding to the field
envelope and $\hat{A}^\dag$ its hermitic conjugated. $\hat{A}$ and
$\hat{A}^\dag$ satisfy the bosonic \emph{equal-space} commutation
rule:
\begin{equation}\label{commut1}
\left[\hat{A}(z,t),\hat{A}^\dag(z,t')\right]=\hbar\omega_0
\delta(t-t').
\end{equation}
The normalization constant $\hbar\omega_0$ stems from the
normalization chosen for the field envelope. In the quantum
propagation theory, the expectation value of the optical power
flowing through the plane $z=\text{constant}$ at time $t$ is given
by $\langle\hat{A}^\dag(z,t)\hat{A}(z,t)\rangle$. In a one
dimensional system, space and time play a symmetrical role. The
dynamics can be described either in terms of spatial wave-packet
\emph{evolution} (evolution picture) or in terms of temporal
wave-packet \emph{propagation} (propagation picture). The first
picture is the must common in quantum field theory and leads to
equal-time commutation rules between the envelope fields $\hat{A}$
and $\hat{A}^\dag$. In this section however, we chose to work in
the propagation picture (which is usual in nonlinear optics) in
order to get a closer correspondence between quantum (\ref{QNLS})
and classical (\ref{NLS}) equations. In this case equal-space
commutation rules (\ref{commut1}) must be imposed to $\hat{A}$ and
$\hat{A}^\dag$ \cite{Blow90}. (In contrast, the evolution picture
is used in the Appendix~\ref{app:A} to derive the stochastic
equations of Sec.~\ref{sec:3}.)

In the cw regime, the basics of the quantum theory of S-MI can be
understood by performing a first-order perturbation analysis of
the steady state solution of Eq.~(\ref{QNLS}). Because the pump
field contains a large number of photons one can treat it as a
classical coherent wave. Using this approximation the steady-state
solution is just the classical one:
$A_{st}(z)=\sqrt{P_0}\exp{(i\gamma P_0 z)}$. The disturbed field
can be written as:
\begin{equation}\label{perturb}
\hat{A}(z,t)=A_{st}(z)+\hat{a}(z,t).
\end{equation}
The disturbance operator $\hat{a}(z,t)$ is defined by
Eq.~(\ref{perturb}). It satisfies the same commutation rule as
$\hat{A}(z,t)$:$\left[\hat{a}(z,t),\hat{a}^\dag(z,t')\right]=
\hbar\omega_0 \delta(t-t')$. Injecting the ansatz (\ref{perturb})
into Eq.~(\ref{QNLS}) one obtains a propagation equation for the
disturbance $\hat{a}(z,t)$. Supposing that the disturbance is
small (non depleted pump approximation) this equation can be
linearized and solved analytically in the Fourier domain
\cite{Potasek87}.

Using this method one finds that the quantum and classical theory
predict the same frequency dependence of the parametric gain. In
contrast the quantum and classical theory differ in that they do
not predict the same growth law for Stokes and anti-Stokes photon
numbers. One easily finds that the quantum counterparts of
Eqs.~(\ref{S-MIcl}) are
\begin{subequations}\label{S-MIqu}
\begin{eqnarray}
n_s(L) &=& n_s(0)\cosh^2(\gamma P_0 L)\nonumber\\
&\phantom{=}&+[n_a(0)+1] \sinh^2(\gamma P_0L),\\
n_a(L)&=&n_a(0) \cosh^2(\gamma P_0 L)\nonumber\\
&\phantom{=}&+[n_s(0)+1]\sinh^2(\gamma P_0L),
\end{eqnarray}
\end{subequations}
where $n_s$ and $n_a$ stand now for the \textit{expectation
values} of the photon number operators:
$n_i=\langle\hat{n}_i\rangle$, $i=s,a$. These equations show that
S-MI can be observed even in absence of any classical input noise.
Setting $n_s(0)=n_a(0)=0$ in Eqs.~(\ref{S-MIqu}) gives the number
of Stokes and anti-Stokes photons produced by the sole action of
vacuum fluctuations:
\begin{equation}\label{S-MIvf}
n_s(L)=n_a(L)=\sinh^2(\gamma P_0L).
\end{equation}
Eqs.~(\ref{S-MIqu}) hold only for perfectly phase-matched photons
at frequencies $\omega_s$ and $\omega_a$ and uncorrelated initial
noise but can be easily generalized to get round these
restrictions.

Although Eqs.~(\ref{S-MIqu}) and (\ref{S-MIvf}) give a good
insight into the physics of vacuum-fluctuations induced S-MI and
photon pairs generation, they are not suitable for quantitative
predictions in the pulsed regime. This is because the effective
value of the pump power $P_0$ depends on pump pulse shape,
duration, and spectral width. Furthermore the energy spectral
density of Stokes and anti-Stokes waves deduced from
(\ref{S-MIvf}) is highly dependent on the precise definition of
modes. This will be discussed in Sec.~\ref{sec:4b}. In the next
sections we will show how to get around these difficulties by
introducing the SNLSE and solving it numerically.

\section{\label{sec:3}Scalar and vector modulation
\protect\\instabilities: Stochastic equations}

One can go beyond the perturbation analysis of S-MI by solving the
QNLSE (\ref{QNLS}) numerically. Such a plan could seem cumbersome
because Eq.~(\ref{QNLS}) is a field-operator equation. The problem
can be bypassed by converting the operator equation (\ref{QNLS})
into c-number equations. This can be performed by choosing a
representation for the electromagnetic field. In this article, we
will use the positive $P$-representation ($P^{(+)}$) introduced by
Drummond and Gardiner \cite{Drummond80}. The c-number equations
obtained in this way are not standard deterministic partial
derivative equations but stochastic (Langevin-type) ones.

Using the $P^{(+)}$ representation, it can be shown
\cite{Drummond87,Carter87,Kennedy88a} that the QNLSE (\ref{QNLS})
is equivalent to the following set of stochastic equations:
\begin{widetext}
\begin{subequations}\label{SNLS}
\begin{eqnarray}
\frac{\partial A}{\partial z}+\frac{1}{v_g}\frac{\partial A}{\partial t}&=&
-i\frac{\beta_2}{2}\frac{\partial^2A}{\partial t^2}
+i\gamma [A^\dag(z,t) A(z,t)]A(z,t)+\sqrt{i\gamma \hbar \omega_0}~\zeta_1(z,t)~A(z,t),\\
\frac{\partial A^\dag}{\partial z}+\frac{1}{v_g}\frac{\partial
A^\dag}{\partial t}&=&
+i\frac{\beta_2}{2}\frac{\partial^2A^\dag}{\partial t^2}-i\gamma
[A^\dag(z,t) A(z,t)]A^\dag(z,t) +\sqrt{-i\gamma \hbar
\omega_0}~\zeta_2(z,t)~A^\dag(z,t).\phantom{MM}
\end{eqnarray}
\end{subequations}
\end{widetext}
Eqs.~(\ref{SNLS}a) and (\ref{SNLS}b) look like the classical NLSE
(\ref{NLS}) and its complex conjugated, except for the last terms
which accounts for vacuum fluctuations. These last terms contains
two independent zero-mean Gaussian white noise random fields
$\zeta_1(z,t)$ and $\zeta_2(z,t)$ characterized by the following
second order moments:
\begin{equation}\label{moments}
\langle\zeta_k(z,t)\zeta_l(z',t')\rangle=\delta_{kl}\delta(z-z')\delta(t-t'),
\end{equation}
with $(k,l)\in \{1,2\}^2$. Because the random fields $\zeta_1$ and
$\zeta_2$ are not complex conjugated of each other, the envelope
fields $A$ and $A^\dag$ are only complex conjugated "in mean" and
have to be treated as different mathematical objects.

Eqs.~(\ref{SNLS}) can be solved on a computer. The numerical
methods used for this task will be briefly explained in
Sec.~\ref{sec:4a}. Note that solving Eqs.~(\ref{SNLS}) gives a
single realization of the stochastic process. In order to
calculate the expectation value of a quantum observable, a
statistical average on many realizations is required. Thus one
generates a large number of realizations
$(A_{[n]}(z,t),A^\dag_{[n]}(z,t))$, $n=1,\ldots, N$. In order to
calculate the expectation value of a quantum observable one then
carries out a statistical average over the many realizations. For
example, the expectation value of the energy spectral density of a
pulse is given by:
\begin{equation}
\label{SE} S_E(z,\Omega)={1\over N}\sum_{n=1}^N
\tilde{A}_{[n]}^\dag(z,\Omega)\tilde{A}_{[n]}(z,\Omega),
\end{equation}
where $\tilde{X}(z,\Omega)=\int_{-\infty}^\infty X(z,t)~e^{i\Omega
t}~dt$ designates the Fourier transform of the field $X(z,t)$ and
$\Omega$ is the detuning from the pump angular frequency
$\omega_0$.

In practice, a few realizations are enough when the number of
photons per mode is high. However in a regime dominated by vacuum
fluctuations hundreds of realizations are typically required to
get precise values. A comparison between Stokes and anti-Stokes
photon production predicted by the SNLSE (\ref{SNLS}) and the
analytical formulas (\ref{S-MIqu}) will be presented in sections
\ref{sec:4b} and \ref{sec:4c}.

The V-MI occurs in birefringent fibers. These fibers are
characterized by different propagation constants $\beta_{0x}$ and
$\beta_{0y}$ and different group velocities $v_{gx}$ and $v_{gy}$
for the $x$- and $y$-axis polarized modes. The numerical study of
vacuum fluctuations induced V-MI requires an extension of
Eqs.~(\ref{SNLS}) that takes into account the phase mismatch
parameter $\Delta\beta_0=\beta_{0x}-\beta_{0y}$ as well as the
group-velocity mismatch $\Delta\beta_1=1/v_{gx}-1/v_{gy}$. Such
extensions have been established in earlier works for the
low-birefringence \cite{Kennedy88b} and the hight-birefringence
\cite{Kennedy91} limits. We generalized these results for an
\emph{arbitrary} level of birefringence and obtained the following
stochastic coupled nonlinear Schr\"odinger equations:
\begin{widetext}
\begin{subequations}\label{SCNLS}
\begin{eqnarray}
\frac{\partial A_x}{\partial z}+\frac{1}{v_{gx}}\frac{\partial A_x}{\partial t}&=&
-i\frac{\beta_2}{2}\frac{\partial^2A_x}{\partial t^2}
+i\gamma \left[A_x^\dag A_x+(1-B)A_y^\dag A_y\right] A_x
+i\gamma B (A_y)^2A_x^\dag e^{-2i\Delta\beta_0z}\nonumber\\
&\phantom{=}&+\sqrt{i\gamma \hbar \omega_0}\left[\zeta_1 A_x+\sqrt{B} \zeta_3A_ye^{-i\Delta\beta_0 z}\right],\\
\frac{\partial A_x^\dag}{\partial z}+\frac{1}{v_{gx}}\frac{\partial A_x^\dag}{\partial t}&=&
+i\frac{\beta_2}{2}\frac{\partial^2A_x^\dag}{\partial t^2}
-i\gamma \left[A_x^\dag A_x+(1-B)A_y^\dag A_y\right] A_x^\dag
-i\gamma B (A_y^\dag)^2A_x e^{+2i\Delta\beta_0z}\nonumber\\
&\phantom{=}&+\sqrt{-i\gamma \hbar \omega_0}\left[\zeta_2 A_x^\dag+\sqrt{B} \zeta_4A_y^\dag e^{+i\Delta\beta_0 z}\right],\\
\frac{\partial A_y}{\partial z}+\frac{1}{v_{gy}}\frac{\partial A_y}{\partial t}&=&
-i\frac{\beta_2}{2}\frac{\partial^2A_y}{\partial t^2}
+i\gamma \left[A_y^\dag A_y+(1-B)A_x^\dag A_x\right] A_y
+i\gamma B (A_x)^2A_y^\dag e^{+2i\Delta\beta_0z}\nonumber\\
&\phantom{=}&+\sqrt{i\gamma \hbar \omega_0}\left[\zeta_1 A_y-\sqrt{B} \zeta_3A_xe^{+i\Delta\beta_0 z}\right],\\
\frac{\partial A_y^\dag}{\partial z}+\frac{1}{v_{gy}}\frac{\partial A_y^\dag}{\partial t}&=&
+i\frac{\beta_2}{2}\frac{\partial^2A_y^\dag}{\partial t^2}
-i\gamma \left[A_y^\dag A_y+(1-B)A_x^\dag A_x\right] A_y^\dag
-i\gamma B (A_x^\dag)^2A_y e^{-2i\Delta\beta_0z}\nonumber\\
&\phantom{=}&+\sqrt{-i\gamma \hbar \omega_0}\left[\zeta_2 A_y^\dag-\sqrt{B} \zeta_4A_x^\dag e^{-i\Delta\beta_0 z}\right],
\end{eqnarray}
\end{subequations}
\end{widetext}
where $(A_x, A_x^\dag)$ and $(A_y, A_y^\dag)$ are stochastic
envelope fields associated to the $x$- and $y$-axis modes
respectively, and $B=\chi_{xyyx}/\chi_{xxxx}$ is a parameter that
measures the strength of the nonlinear coupling between the $x$
and $y$ components. Its value lies between 0 and 1 and depends on
the nonlinearity mechanism. For silica fiber, we can set $B=1/3$
because the Kerr non linearity has principally an electronic
origin. Four independent gaussian random fields $\zeta_k(z,t)$ are
needed to reproduce the effect of vacuum fluctuations. They are
characterized by the second order moments (\ref{moments}), as in
the scalar case, with $(k,l)\in \{1,2,3,4\}^2$. The demonstration
of this set of equations is outlined in the Appendix~\ref{app:A}.

Note that, in contrast to the scalar case, a perturbation analysis
does not lead to simple analytical formulas for
vacuum-fluctuations induced V-MI. For this reason
Eqs.~(\ref{SCNLS}) constitute a valuable theoretical tool.

\section{\label{sec:4}Numerical integration of the SNLSE}

\subsection{\label{sec:4a}Sample Spectra}

We have developed a method for integrating numerically the SNLSE
in the case where the pump is represented by a pulse of finite
duration. Our method is based on the split-step Fourier (SSF)
method \cite{Agrawal95}. We have had to generalize the method in
two ways. First of all the stochastic noise is modeled by
including a noise term at each propagation step. Second in the
case of the V-MI we must alternate not only between the time and
Fourier domain, but also between the linear and circular
polarization basis. Switching from time to Fourier domain is the
basics of the SSF method: it permits to handle the
time-derivatives in a simple way. Similarly, it turns out that,
whereas the terms with time-derivatives are easier to handle in
the linear polarization basis, the $\gamma$-dependent terms are
better managed in the circular polarization basis.

The quantum noise in the SNLSE (\ref{SCNLS}) contains four
independent real zero-mean Gaussian white noise functions
$\zeta_{k}(z,t)$ characterized by Eq.~(\ref{moments}). In the
numerical method, time and space are discretized with respective
discretization steps $\tau$ and $h$. So each family of noise
functions $\zeta_{k}(z,t)$ becomes a finite number of random
variables $\zeta_{k}[i_{z},i_{t}]$ chosen according to a zero-mean
Gaussian law of variance $\frac{1}{h\tau}$. The variance value is
imposed by the normalization condition (\ref{moments}). The
matrixes $\zeta_{k}[i_{z},i_{t}]$ ($k=1,...,4$) define the
stochastic path of each realization. In contrast, when we will
study the effect of classical noise we will add it once, at the
beginning of the pulse propagation, to the spectral distribution
of the signal $\tilde{A}(0,\Omega)$.

As we have previously indicated only the expectation values of
observables have a physical meaning in the stochastic equations.
From a numerical point of view this means that one must average
the calculated quantities over several realizations of the
stochastic path, and/or the classical input noise. Usually,
averaging over a hundred of realizations gives an uncertainty on
the numerical results less than $1$~dB in the non-zero gain
frequency-range. Finally we note that including the stochastic
terms do not increase significantly the numerical complexity of a
single realization.

\begin{figure}[t]
\includegraphics[width=8.2cm,height=13.5cm]{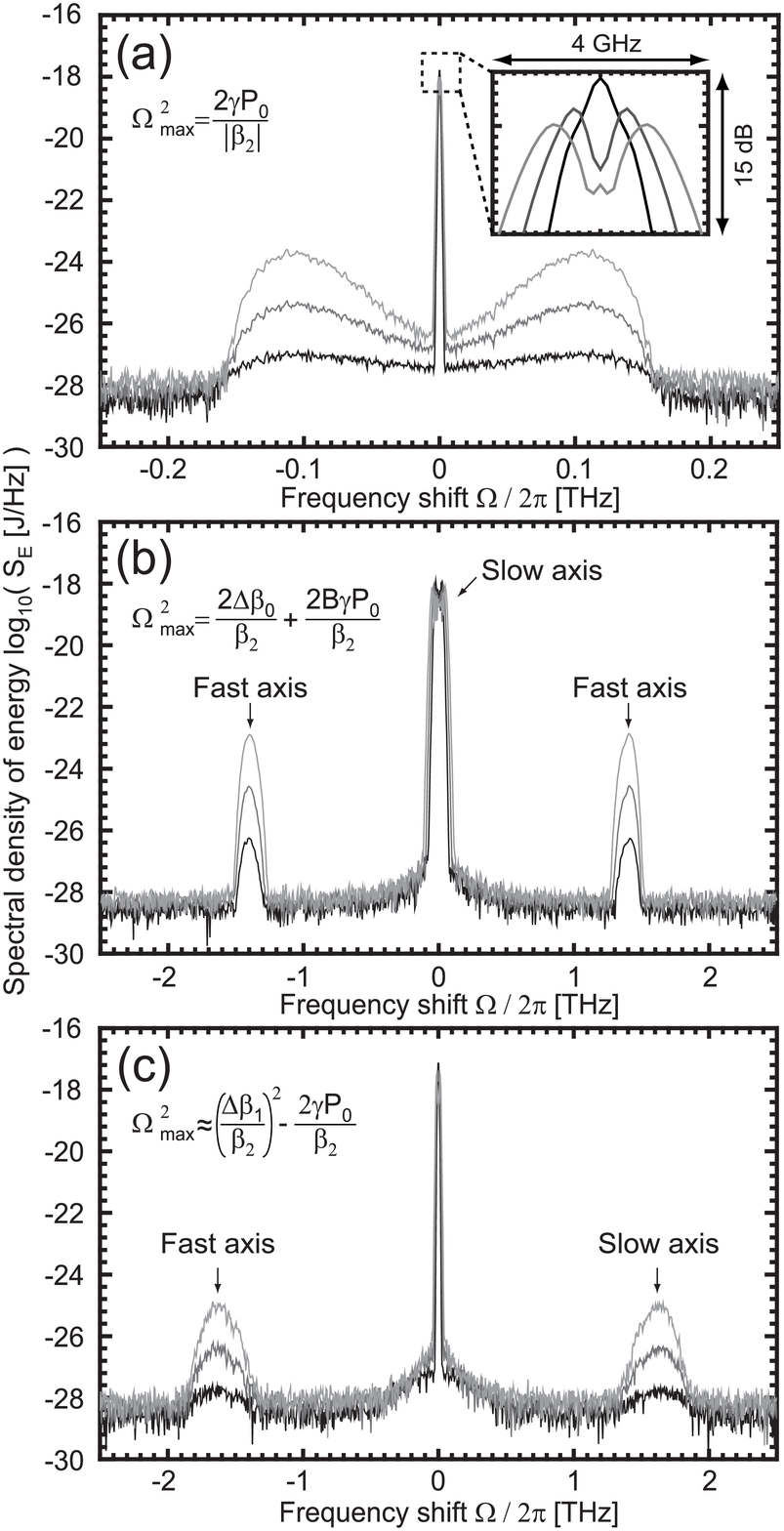}\\%
\caption{MI spectra for various birefringence regime. Simulation parameters are listed %
in Tab.~\ref{table1}. (a) Spontaneous %
S-MI in perfectly isotropic fiber in the anomalous %
dispersion regime. Black, dark gray and light gray curves correspond to %
a propagation length $L$ = 500, 1000, 1500~m, respectively. The inset exhibits %
the pump spectral broadening due to the self-phase-modulation (SPM) effect. %
(b) Spontaneous %
V-MI in a slightly birefringent fiber in the normal dispersion regime. %
The pump is polarized along the slow axis, Stokes and anti-Stokes photons appear %
on the orthogonal axis. Black, dark gray and light gray curves correspond to %
a propagation length $L$ = 16, 24, 32~m, respectively. (c) Spontaneous V-MI in %
a strongly birefringent fiber in the normal dispersion regime. The pump polarization %
axis makes an angle of 45 degrees with the slow axis. Stokes (anti-Stokes) photons appear %
on the slow (fast) axis. Black, dark gray and light %
gray curves correspond to a propagation length $L$ = 10, 20, 30~m,
respectively.} \label{Fig1}
\end{figure}

Some sample spectra obtained using our algorithm are presented in
Fig.~\ref{Fig1} in the cases of S-MI (Fig.~\ref{Fig1}a), low
birefringence V-MI (Fig.~\ref{Fig1}b), and high birefringence V-MI
(Fig.~\ref{Fig1}c). The energy spectral density is plotted versus
the frequency detuning from the pump. The physical parameters used
in these simulations are listed in Table~\ref{table1}. In every
simulation, the pump wave has been supposed to be an unchirped
linearly polarized Gaussian pulse with peak power $P_0$ and
full-width-half-maximum duration $T_{\text{FWHM}}$. No classical
noise has been added, and an average over 50 realizations of the
stochastic process has been performed.
\begin{table}[h]
\begin{ruledtabular}
\begin{tabular}{c c|c c c}
& &Fig.~\ref{Fig1}a & Fig.~\ref{Fig1}b & Fig.~\ref{Fig1}c \\\hline
$\lambda_0$ &[nm] & 1550 & 1550 & 1064 \\
$\beta_2$ &[ps$^2$km$^{-1}$] & -17 & +60 & +30 \\
$\gamma$ &[W$^{-1}$km$^{-1}$]& 2 & 2 & 2 \\
$T_{\text{FWHM}}$ &[ns]& 1 & 0.1 & 0.2 \\
$P_0$ &[W]& 2 & 400 & 300 \\
$\Delta\beta_0$ &[m$^{-1}$]& 0 & 2.09 & 628.31 \\
$\Delta\beta_1$ &[fs m$^{-1}$] & 0 & 1.72 & 354.91 \\
$\theta$\footnote{Angle between the slow axis and the pump polarization axis.}%
& [degree] & 0 & 0 & 45 \\
\end{tabular}
\end{ruledtabular}
\caption{Simulation parameters for Fig.~\ref{Fig1}a-c.} \label{table1}
\end{table}
The frequency detuning of the side bands agrees with linear
perturbation theory. (In order to make the comparison easier we
have indicated in Fig.~\ref{Fig1}a-c the angular frequency shifts
$\Omega_{\text{max}}$ at which maximum gain is expected on the
basis of the linear perturbation theory.)

Moreover the SNLSE predict quantitatively the effect of vacuum
fluctuations on the evolution of the energy spectral density of
the electromagnetic field. In Sec.~\ref{sec:4b} we will show that
this evolution is also in very good agreement with linear
perturbation theory when the number of Stokes and anti-Stokes
photons generated is small enough. When the side bands are
well-developed, the perturbation theory fails to predict correct
values of $S_E$. In contrast, the SNLSE algorithm still gives
accurate results. In this limit numerical results can be easily
confronted to experimental data. In \cite{Amans}, we report an
experiment on high birefringence spontaneous V-MI in the anomalous
dispersion regime that shows that the theoretical spectra from the
SNLSE model tally with the experimental ones.

It is interesting to note that the curves of Fig.~1a look more
noisy than those of Fig.~1b although we have performed the same
number of stochastic realizations in both cases. This is because
the number of realizations needed to achieve a given precision on
the expectation value of a quantum operator is a function of the
relative value of the spatial step $h$ and the typical distance
over which the nonlinear effects act; both are different in the
simulations of Fig.~\ref{Fig1}a and Fig.~\ref{Fig1}b. In practice,
the smaller the spatial step, the fewer the number of realizations
needed to achieve a given precision on expectation values. In
Fig.~\ref{Fig1}, 50 realizations are enough to estimate $S_E$ with
an accuracy of about $1.5$~dB.
We also point out that the noise level visible at non
phase-matched frequencies has no physical meaning. It can be
lowered by averaging over a higher number of realizations.
However, the number of realizations needed to achieve an accurate
estimation of $S_E$ in the non phase-matched part of the spectrum
is usually very high. When tractable, a linear perturbation
analysis will be less time-consuming.

\begin{figure}[h]
\includegraphics[width=8.2cm]{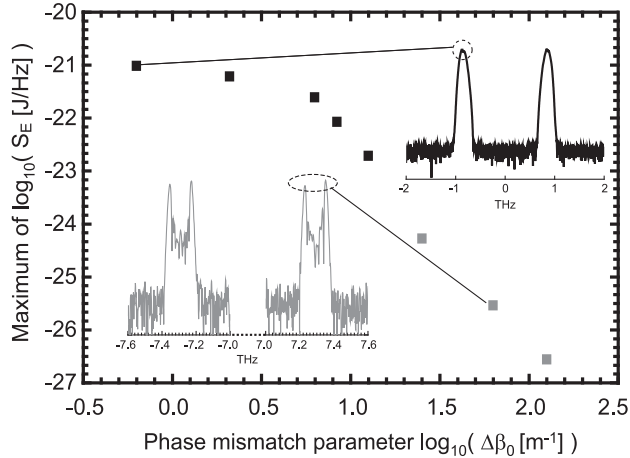}\\%
\caption{The figure illustrates the effect of an increasing
birefringence on the weak-birefringence V-MI phenomenology. The
parameters of the simulations are: $\lambda_0=1550$~nm,
$\beta_2=$~60~ps$^{2}$~km$^{-1}$, $\gamma=2$~W$^{-1}$~km$^{-1}$,
$T_{\text{FWHM}}=100$~ps, $P_0=$~400~W, and $L=$~ 40~m. The pump
wave is polarized along the slow axis. The fiber beat-length was
varied from 10~m to 5~cm. The values of $\Delta\beta_0$ and
$\Delta\beta_1$ where deduced from Eq.~(\ref{biref}). The plot
shows the maximum value of $S_E$ in the side bands as a function
of $\Delta\beta_0$. Black squares (gray squares) correspond to a
single (double) peak structure (see the text).} \label{Fig2}
\end{figure}

Because our algorithm permits to investigate intermediate
birefringence, we have also studied the effect of group velocity
mismatch on the transition from low to high birefringence limits.
To our knowledge, this transition has never been fully
investigated before. Fig.~\ref{Fig2} shows the results of
simulations with the same parameters as in Fig.~\ref{Fig1}b except
that the propagation length has been set to $L=40$~m and that the
fiber beat length $L_B$ was varied from 10~m to 5~cm. The value of
the maximal energy spectral density $S_E$ in the side bands is
plotted versus the phase mismatch parameter. By lowering $L_B$, we
increase the value of the phase mismatch $\Delta\beta_0$ and the
group-velocity mismatch $\Delta\beta_1$ according to the relations
\begin{equation}\label{biref}
\Delta\beta_0=\frac{2\pi}{L_B}, \quad
\Delta\beta_1=\frac{\lambda_0}{c L_B},
\end{equation}
where $c$ is the vacuum speed of light. When the birefringence
increases, the side bands move away from the pump spectrum and
their amplitude decreases. Subsequently the side bands acquire a
double peak structure (see Fig.~\ref{Fig2}). This behavior is due
to the walk-off of the produced Stokes and anti-Stokes
photons. One easily shows that $1/v_{ga}%
-1/v_{gs}=\sqrt{8\beta_2\Delta\beta_0}$, where $v_{gs}$ and
$v_{ga}$ are the Stokes and anti-Stokes photons group velocities,
respectively. Applying this formula to our simulation and taking
$\Delta\beta_0=10$~m$^{-1}$, one sees that the Stokes and
anti-Stokes photons have walked 87.6 ps away while their FWHM
duration is 100 ps. Stokes and anti-Stokes walk-off limits the
coherent exponential amplification of quantum noise. The typical
length scale over which the side bands growth takes place is given
by $T_{\text{FWHM}}/\sqrt{8\beta_2\Delta\beta_0}$. We point out
that this analysis also hold for a cw-pump: The coherent
amplification of the side bands stops when the walk-off of Stokes
and anti-Stokes photons exceeds the coherence length of the pump.
However, in the cw-case, Stokes and anti-Stokes photons generated
in the first coherence length act as an input noise that will be
amplified in the following coherence length. The process is
reproduced as many times as the number of coherence lengths in the
propagation distance. One usually argues \cite{Menyuk87,Agrawal95}
that the weak birefringence phenomenology disappears because the
coherent-coupling terms in Eqs.~(\ref{SCNLS}) (those containing
the factor $\exp[\pm (2)i\Delta\beta_0z]$) average to zero when
$\Delta\beta_0$ is high. This statement is equivalent to saying
that $\Omega_{\text{max}}$ tends to infinity. Our analysis shows
however that walk-off has an even stronger effect.

Until now we have not yet demonstrated that modeling
vacuum-fluctuations induced MI using SNLSE predicts the correct
values of $S_E$. In subsections \ref{sec:4b} we compare the
absolute values of the energy spectral density at the maximum gain
obtained using our program and the linear perturbation analysis.
Having clarified in this way the interpretation of the results of
our numerical simulations we turn to a detailed comparison of the
effect of classical and quantum noises. That is we compare the
effects of classical noise in the initial conditions and the
quantum noise added at each step of the integration.

\subsection{\label{sec:4b}Comparing numerical integration and
linear perturbation theory}
In our numerical simulations we have taken the pump laser to be a
Gaussian pulse without chirp. Its instantaneous power and energy
spectral density can be written
\begin{subequations}\label{InputSignal}
\begin{eqnarray}
P(t)&=&P_{0}\exp(-\frac{t^{2}}{2\sigma_{t}^{2}}),\\
S_E(\Omega)&=&\frac{P_0}{2\sigma_\omega^2}\exp(-\frac{\Omega^{2}}{2\sigma_{\omega}^{2}})
\end{eqnarray}
\end{subequations}
with $\sigma_t\sigma_\omega=\frac{1}{2}$. The numerical
integration of the SNLSE provides us with the spectral density of
energy $S_E(L,\Omega)$ at the end of the fiber, see
Eq.~(\ref{SE}). On the other hand the linear perturbation theory
is based on small perturbation analysis around a continuous
monochromatic pump. We would like to compare quantitatively the
predictions of these two approaches.

For definiteness we carry out this
comparison in the case of scalar MI. We shall focus our
investigation on the intensity of the sidebands at the peak of the
MI gain ($\Omega=\Omega_{\text{max}}$) in the two approaches when
we modify the propagation length and as we modify the duration
$\sigma_t$ of the pulse.

\begin{figure}[h]
\centerline{\includegraphics[width=8.2cm]{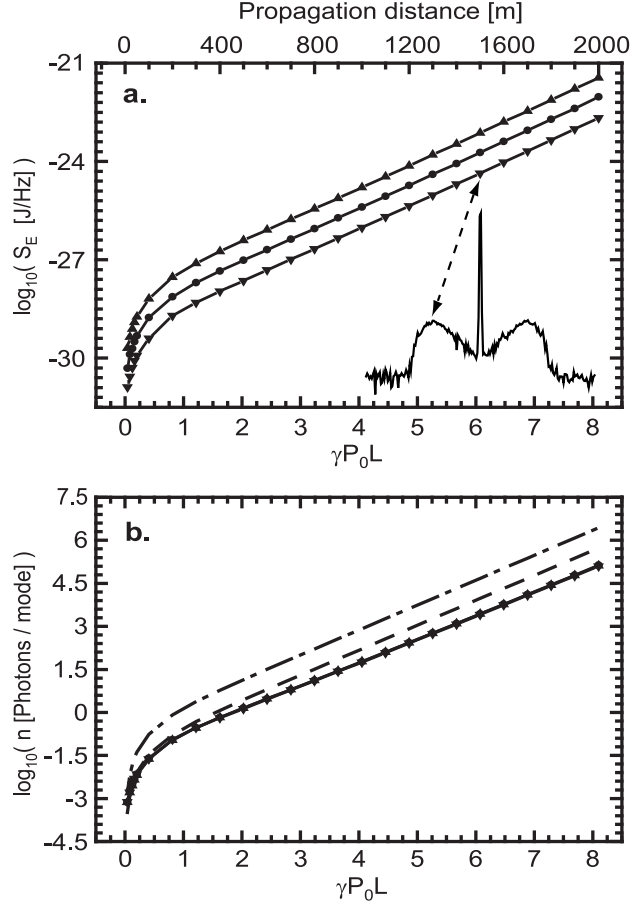}}
\caption{Values of the number of photons created at the maximum
gain frequency obtained by integrating numerically the SNLSE as a
function of propagation length $L$. In order to keep the maximum
gain frequency constant we have kept the peak power $P_0$ constant
in each figure. The horizontal axis is given in dimensionless
units of $\gamma P_0 L$ where $P_0$ is either the power of the
continuous pump wave in the linear approximation, or the peak
power of the Gaussian pump pulse. The top panel is plotted in the
density of energy representation whereas the bottom panel is
plotted in the number of photons-per-mode representation using the
rescaling of Eq. (\ref{nOmega}). In both panels, the up-triangles,
the circles and the down-triangles correspond to FWHM durations
respectively equal to 4~ns, 1~ns and 0.25~ns. Note that these
three curves coincide perfectly in panel (b), thereby showing the
relevance of the rescaling (\ref{nOmega}). In the bottom panel,
the dash-dotted line results from the analytic solution
Eq.~(\ref{S-MIvf}). The dashed line corresponds to
Eq.~(\ref{S-MIvf}) convoluted with the pump shape then rescaled
according to Eq.~(\ref{nOmega}), see the text. All other
parameters are identical to those used in Fig.~\ref{Fig1}a.}
\label{Fig3}
\end{figure}

The linear perturbation theory is based on a continuous
monochromatic pump. For this reason the theory predicts a {\em
rate} of photon production per unit time. This suggests that if
one takes the pump to be a pulse localized in time, the number of
photons produced should be proportional to the pulse duration, all
other parameters being kept constant. Simulations based on SNLSE
confirm this phenomenology. This is illustrated in
Fig.~\ref{Fig3}a, where the energy spectral density at
$\Omega_{\text{max}}$ is plotted as a function of $\gamma P_0 L$
for three different pulse durations. Note however that the above
argument is valid for square pulses but is not very satisfactory
for Gaussian ones. A better understanding of the origin of this
scaling can be obtained by making appeal to the notion of mode and
of Heisenberg box.

In order to introduce the notion of mode, recall that a temporal
signal can be represented by a distribution in the time-frequency
plane. But because of the time-frequency uncertainty relations, a
point in this plane has no physical meaning. This problem is well
known in signal processing where one usually thinks in terms of
local time-frequency decompositions, using windowed Fourier
transforms (WFT) or wavelet transforms \cite{Mallat99}. Such a
local time-frequency decomposition allows one to decompose a
signal into orthogonal local functions, called \emph{modes}.
These can be depicted as surface elements in the time-frequency
plane. Fourier-transform limited pulses, such as our pump pulse
(\ref{InputSignal}), are represented by Gaussian distributions on
the time-frequency plane, called the Wigner-Ville distributions.
This distribution is very similar to the Wigner distribution used
in quantum optics to represent a quantum state of light. In
particular two different Fourier-limited Gaussian pulses
sufficiently different in time or central frequency can be
considered as quasi-orthogonal modes. A set of quasi non
overlapping Gaussian Wigner-Ville distributions can be taken as a
base for the time-frequency decomposition of the field. One can
visualize this modal base by imagining that the time-frequency
plane is paved with adjacent elementary surface elements, called
Heisenberg boxes, roughly representing the area of Gaussian quasi
non overlapping Wigner-Ville distributions. The precise area of
the Heisenberg boxes is a matter of taste depending on how strong
orthogonality is required. A usual convention is to take this area
equal to $\sigma_t\times \sigma_\omega=1/2$.

This set of (quasi) mode is convenient for our problem because,
during the pulse propagation in the fiber, the uncertainty on the
creation time of a photon is defined by the variance
$\sigma_{t}^{2}$ of the pump pulse, and implies an uncertainty on
its frequency defined by the variance $\sigma_{\omega}^{2}$. More
precisely, in the case of S-MI, the pump pulse (\ref{InputSignal})
produces photons that occupy single Heisenberg boxes located at
the same time as the pump but at angular frequencies
$\omega_0\pm\Omega$ ($0<\Omega<\sqrt{2}\Omega_{\text{max}}$).
Formulas (\ref{S-MIqu}) and (\ref{S-MIvf}) of Sec.~\ref{sec:2b}
thus give the number of photons created in these Stokes and
anti-Stokes modes.

Our numerical simulations provide us with the spectral
distribution of energy $S_{E}$. Hence we need to reexpress this as
the number $n$ of photons produced per mode of duration $\sigma_t$
and spectral width $\sigma_\omega$:
\begin{equation}
n(\Omega)=\frac{S_{E}(\Omega)}
{\hbar\omega_{0}}\times\sigma_{\omega}=\frac{S_{E} (\Omega)}
{\hbar\omega_{0}} \frac{\sqrt{2\ln2}}{T_{\text{FWMH}}}.
\label{nOmega}
\end{equation}
Fig.~\ref{Fig3}b shows the effect of scaling the spectra according
to Eq.~(\ref{nOmega}). When expressed in terms of number of
photons per modes the three curves of Fig.~\ref{Fig3}a
(corresponding to three different pump durations) come down to a
single one in Fig.~\ref{Fig3}b (continuous line). This shows that
the notion of mode helps in interpreting the results of the
numerical integration. The number of produced photons does not
depend of the pump duration: The pump duration just alters their
time-frequency characteristics. The dash-dotted curve in
Fig.~\ref{Fig3}b corresponds to a direct application of
Eq.~(\ref{S-MIvf}). A discrepancy with the simulations based on
the SNLSE can be noted. It is simply due to the fact that our
choice of size of Heisenberg boxes, hence the normalization factor
in Eq.~(\ref{nOmega}), is somewhat arbitrary. By taking the
Heisenberg boxes a bit bigger, one can put the continuous and
dash-dotted curves of Fig~\ref{Fig3}b in superposition. In the
rest of the text, however, we maintain the normalization relation
Eq.~(\ref{nOmega}) for clarity.

There is another way to compare the results of SNLSE simulations
to the linear perturbation theory that avoids the concept of
modes. One first computes the power spectral density given by the
quantum perturbation analysis on a monochromatic pump wave. Then
one convolutes this with the Gaussian spectral distribution of the
real pump pulse. This procedure gives a good approximation of the
energy spectral density generated by the Gaussian pump. The dashed
curve of Fig.~\ref{Fig3}b corresponds to the peak energy spectral
density computed by this method and rescaled according to
Eq.~(\ref{nOmega}) in order to be independent of the pulse
duration. The agreement with the continuous curve is now much
better. The origin of the small difference in slope in the
exponential amplification regime still remains unclear. It may be
due to the self-phase-modulation broadening of the pump spectrum,
which is not taken into account by the linear perturbation
analysis.

As a conclusion we obtain a good quantitative agreement between
the simulation based on SNLSE and the linear perturbation
analysis. We have also shown that the only physical quantity that
can be rigorously predicted by the quantum nonlinear propagation
theory is the energy spectral density and that the concept of
mode, although very useful, must be handled with care. Especially
formulas like (\ref{S-MIqu}) and/or (\ref{S-MIvf}) may only be
used as a rough approximation tool because there is no objective
way to define a time-frequency mode.

\subsection{\label{sec:4c}Classical versus Quantum Noise}

From Figs.~\ref{Fig3} it is clear that the spontaneous MI growth
can be divided into two different stages. So long as the number of
particles created by mode is less than one ($n<1$), one is in a
quantum regime dominated by vacuum fluctuations. In contrast, for
$n$ above 1 the modulation instability is amplified exponentially
and quantum effects become negligible.

We now investigate the transition between the quantum and
classical regimes in presence of some \emph{classical} noise.
We have chosen to model this noise by modifying the initial
conditions and adding a white noise to the amplitude of the pump
pulse in the Fourier domain. For definiteness and simplicity, we
continue to focus on scalar modulation instability.

The classical initial noise $\widetilde{N}(\Omega)$ was chosen
according to two criteria: (i) the noise must correspond to a
random fluctuation of the pump amplitude in the time domain and
(ii) the statistic of $\widetilde{N}(\Omega)$ (for each frequency)
must lead to a spectral density of energy
$\langle\widetilde{N}(\Omega)\widetilde{N}^{\ast}(\Omega)\rangle$
constant as a function of the frequency. Several noise definitions
can meet these criteria. We have chosen to study two particular
cases : the pure spectral phase noise
\begin{equation}
\widetilde{N}_{\phi}(\Omega)=\widetilde{A}\exp(i\pi\zeta
(\Omega)), \label{ClassicNoise1}
\end{equation}
and the Gaussian noise
\begin{equation}
\widetilde{N}_{\text{G}}(\Omega)=\frac{\widetilde{A}}{\sqrt{2}}\times
(\zeta_1(\Omega)+i\zeta_2(\Omega)), \label{ClassicNoise2}
\end{equation}
where $\widetilde{A}$ is a real constant and $\zeta_j(\Omega)$ are
independent real zero-mean Gaussian white noise random fields. In
our simulations $\zeta_j (\Omega)$ where discretized and replaced
with random quantities $\zeta_j [i_{\omega}]$ (one for each
discretized frequency) drawn according a zero-mean Gaussian law of
variance 1. We have compared S-MI spectra averaged over 50
realizations, obtained from both classical noises (with quantum
noise set to zero). Both noises lead to equivalent results. The
difference between the spectra is lower than 0.3~dB on the full
spectral span, which is less than the residual averaging noise
(See Fig.\ref{Fig1}a), and therefore negligible. Hereafter, the
classical noise is set according to Eq.~(\ref{ClassicNoise1}). It
corresponds to a white flux of photons without any phase
correlations between each frequency component.

\begin{figure}[h]
\begin{center}
\includegraphics[width=8.2cm]{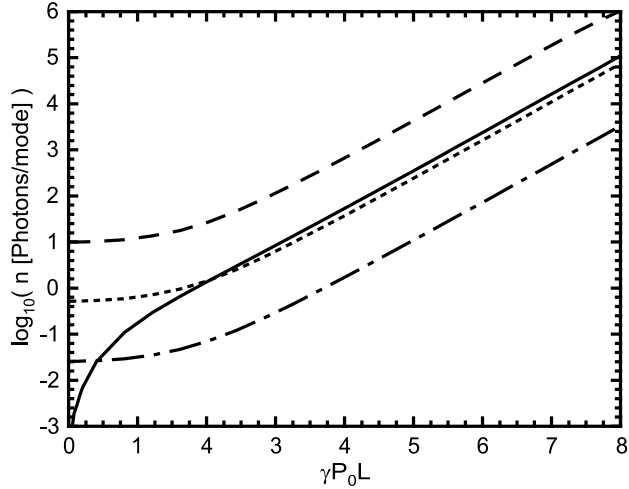}
\caption{Comparison between a MI growing from quantum noise (solid
line) and from purely classical noise. The number of photons in
the Stokes and anti-Stokes modes is plotted as a function of the
propagation length (expressed in dimensionless units $\gamma
P_0L$). The input noises in the classical situation correspond to
the following amount of photons per mode: (i) $1/40$ (dash-dotted
line), (ii) $1/2$ (dotted line), and (iii) 10 (dashed line). The
simulation parameters are those of Fig.~\ref{Fig1}a.}
\label{SansSto}
\end{center}
\end{figure}

In Fig.~\ref{SansSto} we compare the peak intensity of side bands
in the case of vacuum-fluctuations induced MI and the (unphysical)
case of MI induced by the classical noise alone. In the quantum
regime where the number of particles per mode is small the two
approaches differ strongly whereas in the exponential
amplification regime they give similar results, although different
classical noise levels give rise to different final number of
photons.

\begin{figure}[t]
\begin{center}
\includegraphics[width=8.2cm]{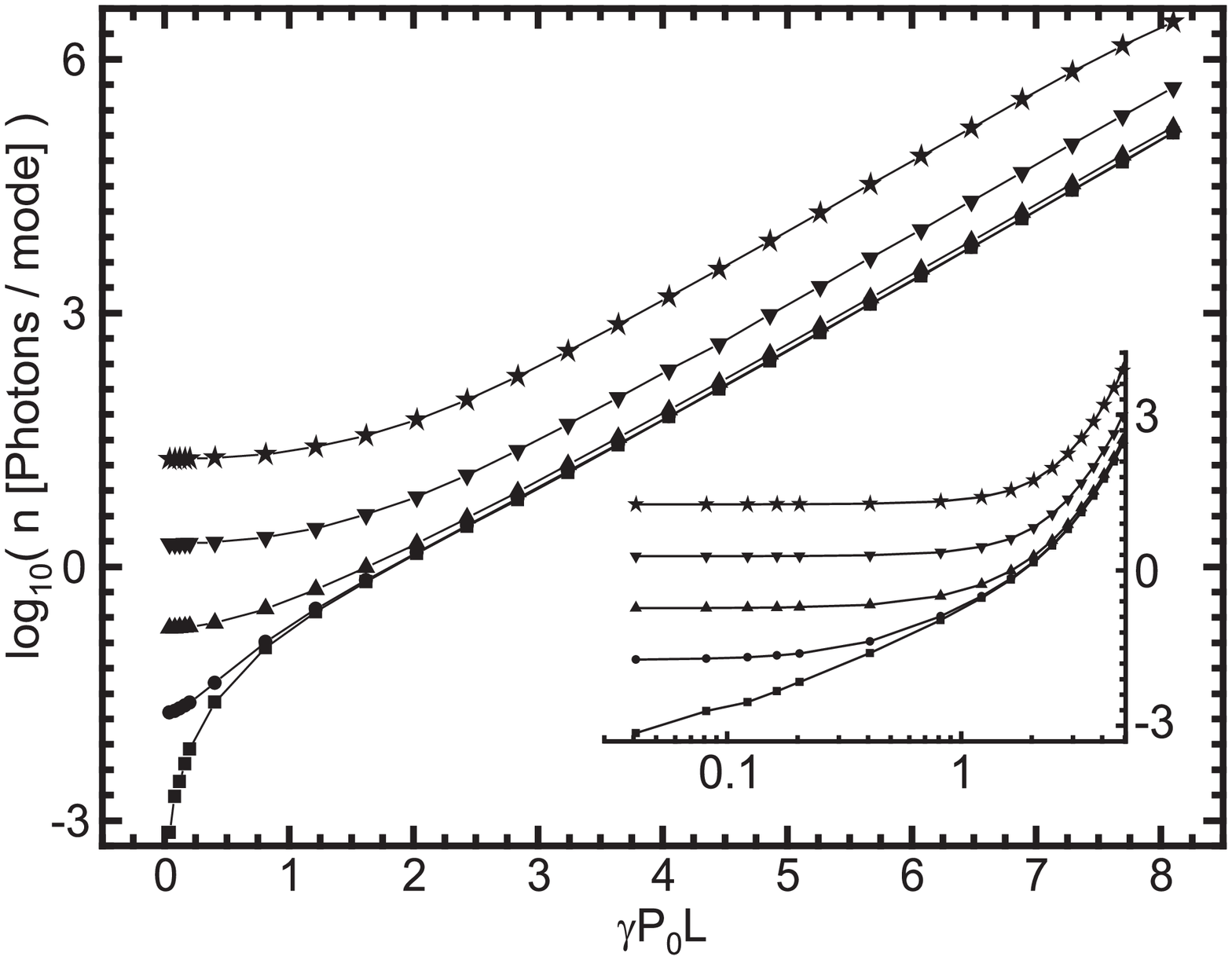}
\caption{Simulated values of the number of photons in the Stokes
and anti-Stokes modes in the realistic case when MI grows from
both quantum and classical noise. The classical noise intensities
are defined in order to correspond to an amount of photons per
GHz: (i) 0.1 (circles), (ii) 1 (up-triangles), (iii) 10
(down-triangles), and (iv) 100 (stars). The squares correspond to
a purely quantum noise. The simulation parameters are those of
Fig.~\ref{Fig1}a. The inset is a zoom corresponding to the the
lowest $\gamma P_{0}L$-parameters. It is drawn in a
logarithmic-scale.} \label{bruit1}
\end{center}
\end{figure}

The simulations of Fig.~\ref{bruit1} take into account both
classical and quantum noises. They illustrate the realistic
situation when both noises compete for producing modulation
instability. If the order of magnitude of the classical noise is
such that there is less than one photon per mode, the quantum
noise dominates and the curve is (except for very small values of
the gain) indistinguishable from the case where there is no
classical noise. On the other hand if the number of noise photons
per mode is much larger than one, the classical noise dominates
and the intensity of the sidebands is indistinguishable from
purely classical situations depicted in Fig.~\ref{SansSto}. These
numerical results are consistent with Eqs.~(\ref{S-MIqu}), in
which the term 1 represents the contribution of the vacuum
fluctuation. If $n_s$ or $n_a$ are higher than 1,
Eqs.~(\ref{S-MIqu}) tend to Eqs.~(\ref{S-MIcl}) corresponding to
the classical description.

In summary we have shown that if one considers only the peak
intensity of the sidebands, the quantum origin of the instability
can only be seen in the regime where the number of photons per
mode (produced or initially present) is small whereas when the
number of photons per mode is large the effect of vacuum
fluctuations is indistinguishable from that of classical noise.
Good quantitative agreement between the two approaches in the
exponential amplification regime is obtained when the number of
noise photons is $1/2$ per mode. Note that other quantum effects,
such as two mode squeezing, may be present in the regime where
many photons are produced per mode, but exhibiting them requires
looking at correlations between the two sidebands.

\subsection{Using classical noise to compute the instabilities induced
by vacuum fluctuations}

We can now discuss the well-known trick which consists in
introducing a half photon per mode into a classical simulation
(then removing it) to simulate the spontaneous effects. To see how
this works we compare Eqs.~(\ref{S-MIcl}) which describe the MI
induced by classical noise and Eq.~(\ref{S-MIvf}) which is derived
from the quantum theory and gives the number of photons produced
per mode by the action of vacuum fluctuations alone. Now, if we
introduce the same amount of Stokes and anti-Stokes noise photons
$n_{0}$ in Eqs.~(\ref{S-MIcl}) we find:
\begin{equation}\label{comparison}
\frac{n_a^{Cl}}{2n_{0}}-\frac{1}{2}=\frac{n_s^{Cl}}{2n_{0}}-\frac{1}{2}=\sinh^{2}(\gamma
P_0L)=n_a^{Qu}=n_s^{Qu},
\end{equation}
where $Cl$ and $Qu$ denote respectively the classical approach and
the quantum approach. Taking $n_{0}=\frac{1}{2}$,
Eq.~(\ref{comparison}) shows the agreement between the quantum
predictions (right hand side) and the classical predictions (left
hand side). Note that $n_{0}$ can be any real value (except for
0), hence the spontaneous growth of the number of photons per mode
can be simulated with any number of initial classical photons if
the pump depletion is neglected.

We have compared, using numerical integration, the direct quantum
approach based on the SNLSE (without classical noise) and the
classical approach in which one first integrates the NLSE with
some initial classical noise then rescales the spectra according
to the left side of Eq.~(\ref{comparison}). The discrepancy
between both approaches is measured by the following ratio in dB
scale:
\begin{equation}
\label{eq:ratio}
\eta=10\log_{10}\left(\frac{n^{Qu}}{\frac{n^{Cl}}{2n_{0}}-\frac{1}{2}}\right),
\end{equation}
Applying Eq.~(\ref{eq:ratio}) to the data reported in
Fig.~\ref{SansSto}, one founds that for $\gamma P_0 L$ higher than
0.2, $\eta$ is constant for any classical noise amplitude:
$\eta$=1.9~dB~$\pm$0.3~dB. Note that both noise definitions
Eq.~(\ref{ClassicNoise1}) and Eq.~(\ref{ClassicNoise2}) lead to
the same $\eta$. Moreover this results may be extended to $\gamma
P_0 L$ lower than 0.2 by increasing the number of realizations.
The origin of non-zero value of $\eta$ may come from the mode
definition used in Eq.~(\ref{nOmega}). Indeed the numerical
integration of SNLSE gives the physical spectral density of energy
$S_E$ whereas the calculus trick --- left hand side of
Eq.~(\ref{comparison}) --- leads to values interpreted as a number
of photons per mode which must be rescaled to give $S_E$.

In summary, in the case of pump pulses of finite duration, a full
quantum treatment based on the SNLSE leads to a direct
quantitative prediction whereas the calculus trick only gives a
good approximation whose accuracy is dependent of the modes
definition.

\section{Conclusion}
We generalized to an arbitrary level of birefringence the
stochastic nonlinear Schr\"{o}dinger equations describing the
propagation of pulses through a nonlinear $\chi^{(3)}$ medium with
linear birefringence and group-velocity dispersion, and developed
numerical routines to compute them. Because these stochastic
equations are equivalent to quantum field operator equations, we
used them to compute spontaneous (or vacuum-fluctuations induced)
modulation instability spectra in various birefringence regimes,
including  weak, high but also intermediate birefringence which
has not been studied so far. In particular we showed that the
decline of the number of photons generated by the
weak-birefringence V-MI when the birefringence increases, is
attributable to the increase of the walk-off between Stokes and
anti-Stokes photons, although the weak-birefringence V-MI gain
remains constant. We then investigated the absolute values of the
energy spectral density at the maximum gain in the case of scalar
modulation instabilities induced by vacuum fluctuations. We
obtained good quantitative agreement between the simulation based
on SNLSE and the linear perturbation analysis. Then we have
carried out a detailed comparison of the effect of classical and
quantum noise and shown that the quantum origin of the instability
can only be seen in the regime where the number of photons per
mode produced or initially present is small. Finally we note that
the quantum nonlinear propagation theory predicts the energy
spectral density and that the concept of mode, although very
useful, must be handled with care. The present work forms the
basis for numerical and experimental investigation of
vacuum-fluctuations induced V-MI in regimes which have been little
investigated so far, and we hope to report on this in the near
future \cite{Amans}.

Although we have not developed this aspect in this article the
stochastic equations can also be used to computed intensity
correlations between side bands and predict special quantum
effects like squeezing. For this reason the stochastic equations
(\ref{SCNLS}) are a valuable tool for computing quantum effects in
birefringent nonlinear $\chi^{(3)}$ media, especially optical
fibers. The stochastic model may also be adapted to include Raman
and Brillouin effects (see \cite{Drummond01a} for the scalar
case). Higher order dispersion effects can also be included in a
straightforward way.
\begin{acknowledgments}
We would like to thank Marc Haelterman, Stephane Coen and Eric
Lantz for usefull discussions. We acknowledge financial support
from the Communaut\'{e} Fran\c{c}aise de Belgique under ARC 00/05-
251, and from the IUAP programme of the Belgian government under
Grant No.V-18.
\end{acknowledgments}

\appendix*
\section{\label{app:A} Stochastic coupled
nonlinear Schr\"{o}dinger equations\protect\\}

In order to derive the SNLSE (\ref{SCNLS}), we will proceed in
three stages. First, we will establish the interaction Hamiltonian
that governs field evolution in a lossless, dispersive, and
birefringent fiber. We will only present an heuristic derivation
of this Hamiltonian and put the emphasis on appropriate
approximations. A rigorous derivation requires a discussion of
electromagnetic field quantization in material media
\cite{Hillery84,Drummond90,Blow90,Drummond01a}, which is outside
the scope of this article. Second, we will use this Hamiltonian to
find the Liouville equation describing the evolution of the
density operator of the field and convert it into a Fokker-Planck
equation using the $P^{(+)}$-representation. Finally, we will
establish the connection between the Fokker-Planck equation and
the stochastic equations (\ref{SCNLS}).
\subsection{Linear and Nonlinear Hamiltonians}
In a dispersive birefringent medium the positive-frequency part of
the $s$-polarized electric field component ($s=x,y$) can be
decomposed on monochromatic modes in the following way
\cite{Blow90}:
\begin{equation}\label{A1}
\hat{E}^{(+)}_s(\mathbf{r})=i\int
d\beta\left(\frac{\hbar~\omega_s(\beta)~v_{gs}(\beta)}
{4\pi\epsilon_0 n_s(\beta)cA}\right)^{\frac{1}{2}}\hat{a}_s(\beta)
F(x,y)e^{i\beta z}.
\end{equation}
Eq.~(\ref{A1}) can be seen as defining $\hat{a}_s(\beta)$. The
operators $\hat{a}_s(\beta)$ and its hermitic conjugated
$\hat{a}_s^\dag(\beta)$ are, respectively, the annihilation and
creation operators for a $s$-polarized photon propagating in the
fiber with a propagation constant $\beta$ and having an angular
frequency $\omega_s(\beta)$. According to the canonical
quantization, they satisfy the commutation rule
\begin{equation}\label{A2}
[\hat{a}(\beta),\hat{a}^\dag(\beta')]=\delta(\beta-\beta').
\end{equation}
In Eq.~(\ref{A1}), $n_s(\beta)$ and $v_{gs}(\beta)$ are
respectively the linear index of refraction and group velocity
corresponding to the $s$-polarized monochromatic mode with
frequency $\omega_s(\beta)$, and
\begin{equation}\label{A3}
A=\int\int |F(x,y)|^2 dxdy.
\end{equation}
The operator representing the $s$-polarized electric field
component is given by
\begin{equation}
\hat{E}_s(\mathbf{r})=\hat{E}^{(+)}_s(\mathbf{r})+\hat{E}^{(-)}_s(\mathbf{r}),
\end{equation}
where
$\hat{E}^{(-)}_s(\mathbf{r})=[\hat{E}^{(+)}_s(\mathbf{r})]^\dag$
is the negative-frequency part of $\hat{E}_s(\mathbf{r})$.

The total
Hamiltonian $\hat{H}_T$ representing the sum of the vacuum
electromagnetic energy and the dielectric energy stored in the
fiber can be decomposed in a linear part $\hat{H}_L$ and a
nonlinear one $\hat{H}_{NL}$.

The linear part,
\begin{equation}\label{A4}
\hat{H}_{L}=\int d\beta \sum_{s=x,y}\hbar \omega_s(\beta)
\hat{a}_s^\dag(\beta)\hat{a}_s(\beta),
\end{equation}
takes into account the free-field energy and the energy stored
into the dielectric through linear interactions, including the
effects of \textit{linear} dispersion and \textit{linear}
birefringence through the dispersion relations
$\omega_s=\omega_s(\beta)$.

If the field bandwidth is narrow compared to the central angular
frequency $\omega_0$, dispersion can be neglected in the
$\chi^{(3)}$ interactions and the nonlinear part of the
Hamiltonian can be written
\begin{equation}\label{A5}
\hat{H}_{NL}=-\frac{1}{4}\epsilon_0 d \int d^3r\sum_{ijkl}
\chi_{ijkl}E_i^{(-)}E_l^{(-)}E_j^{(+)}E_k^{(+)},
\end{equation}
where $\chi_{ijkl}$ stands for
$\chi_{ijkl}(\omega_0;\omega_0,\omega_0,-\omega_0)$. This
simplified Hamiltonian only takes into account the Kerr effect,
which is the dominant one for quasi-monochromatic fields. Since
the medium is supposed lossless, the $\chi^{(3)}$ tensor has the
full permutation symmetry \cite{Boyd92}. The degeneracy factor
$d=6$ takes this symmetry into account, by counting the number of
way to permute the frequency arguments and the indexes of the
$\chi^{(3)}$ tensor. A further useful approximation is to consider
that the $\chi^{(3)}$ process is isotropic \cite{Svirko98,Boyd92}:
\begin{equation*}
\chi_{ijkl}=\chi_{xxyy}\delta_{ij}\delta_{kl}+\chi_{xyxy}\delta_{ik}\delta_{jl}+\chi_{xyyx}\delta_{il}\delta_{jk}.
\end{equation*}
Since the field bandwidth is supposed narrow, the permutation
symmetry also requires that $\chi_{xxyy}=\chi_{xyxy}$:
\begin{equation}\label{A6}
\chi_{ijkl}=\chi_{xyxy}(\delta_{ij}\delta_{kl}+\delta_{ik}\delta_{jl})+\chi_{xyyx}\delta_{il}
\delta_{jk}.
\end{equation}
Using Eq.~(\ref{A6}), the nonlinear Hamiltonian becomes
\begin{widetext}
\begin{eqnarray}
\hat{H}_{NL}=-\frac{3}{2}\epsilon_0\chi_{xxxx}\int d^3r &\bigg(&%
\sum_{s}
\hat{E}_s^{(-)}\hat{E}_s^{(-)}\hat{E}_s^{(+)}\hat{E}_s^{(+)}
+(1-B) \sum_{s\neq s'}
\hat{E}_s^{(-)}\hat{E}_{s'}^{(-)}\hat{E}_s^{(+)}\hat{E}_{s'}^{(+)} \nonumber\\
&\phantom{\bigg(}&+ B \sum_{s\neq s'}
\hat{E}_s^{(-)}\hat{E}_{s}^{(-)}\hat{E}_{s'}^{(+)}\hat{E}_{s'}^{(+)}\bigg),
\end{eqnarray}
\end{widetext}
where we defined $B=\chi_{xyyx}/\chi_{xxxx}$, and factored out
$\chi_{xxxx}=2\chi_{xyxy}+\chi_{xyyx}$.

Another consequence of the narrow-bandwidth assumption is that the
square-rooted bracket in Eq.~(\ref{A1}) can be taken out of the
integral and one can write
\begin{equation}\label{A8}
\hat{E}^{(+)}_s(\mathbf{r})\approx i \left(\frac{\hbar \omega_0
v_{gs0}}{2\epsilon_0~n_{s0}~c~A}\right)^{1/2}F(x,y)\hat{\psi}_s(z,t)e^{i(\beta_{s0}z-\omega_0t)},
\end{equation}
where
\begin{equation}\label{A9}
\hat{\psi}_s(z,t)=\frac{e^{i(\omega_0t-\beta_{s0}z)}}{\sqrt{2\pi}}\int
d\beta_s~\hat{a}_s(\beta_s)~e^{i\beta_sz}.
\end{equation}
In Eqs.~(\ref{A8}) and (\ref{A9}), $n_{s0}$, $v_{gs0}$, and
$\beta_{s0}$ stand respectively for the index of refraction, the
group-velocity, and the propagation constant at frequency
$\omega_0$ on the $s$-axis. The operator $\hat{\psi}$ is an
envelope operator because fast oscillations in space and time have
factored out. This implies that $\hat{\psi}$ is explicitly
time-dependent in the Schrodinger picture. The operator
$\hat{\psi}_{s}^\dag\hat{\psi}_{s}dz$ represents the number of
$s$-polarized photons in $[z,z+dz]$. One can easily check that
\begin{equation}
[\hat{\psi}_s(z,t),\hat{\psi}_{s'}^\dag(z',t)]=\delta_{ss'}\delta(z-z').
\end{equation}
Using Eq.~(\ref{A8}), $\hat{H}_{NL}$ takes the following simple
form:
\begin{widetext}
\begin{eqnarray}\label{A12}
\hat{H}_{NL}=-\frac{\hbar}{2}\Theta\int &\Big[&
\Big(\hat{\psi}_x^\dag\hat{\psi}_x^\dag\hat{\psi}_x\hat{\psi}_x+
\hat{\psi}_y^\dag\hat{\psi}_y^\dag\hat{\psi}_y\hat{\psi}_y\Big)
+2(1-B) \hat{\psi}_x^\dag\hat{\psi}_y^\dag\hat{\psi}_x\hat{\psi}_y
\nonumber \\%
&\phantom{\Big[}& +B\Big(
(\hat{\psi}_x^\dag)^2\hat{\psi}_{y}^2e^{2i\Delta\beta_0z}
+(\hat{\psi}_y^\dag)^2\hat{\psi}_{x}^2e^{-2i\Delta\beta_0z}\Big)\Big]dz,
\end{eqnarray}
\end{widetext}
where
\begin{equation}
\Theta=\frac{3\hbar\omega_0^2v_{g0}^2\chi_{xxxx}}{4\epsilon_0n_0^2c^2A_{\text{eff}}},
\quad A_{\text{eff}}=\frac{A^2}{\int\int |F(x,y)|^4 dx~dy}.
\end{equation}
Let's note that we have set $n_0\equiv n_{x0}\approx n_{y0}$ and
$v_{g0}\equiv v_{gx0}\approx v_{gy0}$. The linear Hamiltonian
$\hat{H}_L$ can also be expressed in function of the operators
$(\hat{\psi}_s,\hat{\psi}^\dag_s)$, $s=x,y$, by developing
$\omega_s(\beta)$ in a Taylor expansion around $\beta_{s0}$ up to
the second order,
\begin{equation}\label{A13}
\omega_s(\beta)=\omega_0+\omega'_s(\beta-\beta_{s0})
+\frac{\omega''_s}{2}(\beta-\beta_{s0})^2+...,
\end{equation}
where $\omega'_s=\frac{d\omega_s}{d\beta}|_{\beta_0}=v_{gs0}$ and
$\omega''_s=\frac{d^2\omega_s}{d\beta^2}|_{\beta_0}$. Using
Eq.~(\ref{A13}) and inverting Eq.~(\ref{A9}), one finds that
$\hat{H}_L=\hat{U}+\hat{H}'_L$, where
\begin{equation}
\hat{U}=\hbar\omega_0\sum_{s=x,y}\int \hat{\psi}_s^\dag(z)
\hat{\psi}_s(z) dz,
\end{equation}
and
\begin{equation}\label{A16}
\hat{H}'_L=\frac{\hbar}{2}\sum_{s=x,y} \int [i\omega'_s
(\frac{d\hat{\psi}_s^\dag}{dz}\hat{\psi}_s-\hat{\psi}_s^\dag\frac{d\hat{\psi}_s}{dz})
+\omega''_s\frac{d\hat{\psi}_s^\dag}{dz}\frac{d\hat{\psi}_s}{dz}]dz.
\end{equation}
In the Heisenberg picture, the hamiltonian $\hat{U}$ is
responsible of a free oscillation $\exp(-i\omega_0 t)$ of the
fields $\hat{\psi}_s$ ($s=x,y$). This oscillation will cancel out
the explicit oscillation $\exp(i\omega_0 t)$ already present in
the definition (\ref{A9}). For this reason, we prefer to continue
the discussion in the interaction picture:
\begin{eqnarray}
(\hat{\psi}_s)_I&=&\hat{\psi}_s\exp(-i\omega_0 t), \label{A17}\\
\hat{H}_I&=&\hat{H}_T-\hat{U}=\hat{H}'_L+\hat{H}_{NL}.\label{A18}
\end{eqnarray}
To simplify the notations we will drop the $I$ index in later
equations.
\subsection{From Liouville to stochastic equations}
In the quantized theory, the state of the electromagnetic field is
represented by the density operator $\hat{\rho}(t)$. Its
evolution, in the interaction picture, is given by the Liouville
equation
\begin{equation}\label{A19}
i\hbar\frac{d}{dt}\hat{\rho}=[\hat{H},\hat{\rho}],
\end{equation}
where $\hat{H}$ is the Hamiltonian defined by Eq.~(\ref{A18}).
Using Eqs.~(\ref{A16}) and (\ref{A12}), the calculation of the
right-hand side of Eq.~(\ref{A19}) is straightforward, so we do
not write it here explicitly.

In order to obtain stochastic equations from the Liouville
equation (\ref{A19}), we generalized the argument of Drummond and
Gardiner \cite{Drummond80} for monomode fields and their extension
to multimode scalar fields given is \cite{Kennedy88a}. We
introduce the multimode coherent states $|\{\alpha\}\rangle$
defined as the eigenstates of the annihilation operators
$\hat{a}_s(\beta)$
\begin{equation*}
\hat{a}_s(\beta)|\{\alpha\}\rangle=\alpha_s(\beta)|\{\alpha\}\rangle.
\end{equation*}
As a consequence, $|\{\alpha\}\rangle$ are also eigenstates of the
envelope operator (\ref{A17})
\begin{equation*}
\hat{\psi}_s(z)|\{\alpha\}\rangle=\psi_s(z)|\{\alpha\}\rangle,
\end{equation*}
with
\begin{equation*}
\psi_s(z)=\frac{1}{\sqrt{2\pi}}\int
d\beta_s~\alpha_s(\beta_s)~e^{i(\beta_s-\beta_{s0})z}.
\end{equation*}
This suggest the alternative notation
$|\bm{\psi}(z)\rangle\equiv|\{\alpha\}\rangle$, with
$\bm{\psi}(z)=(\psi_x(z),\psi_y(z))$, for the multimode coherent
state. The basic idea of $P^{(+)}$-representation is to expand the
density operator on nondiagonal coherent state projection
operators defined as
\begin{equation}\label{A20}
\hat{\Lambda}(\bm{\Psi}(z))=\frac{|\bm{\psi}(z)\rangle\langle(\bm{\psi^\dag})^*(z)|}
{\langle(\bm{\psi^\dag})^*(z)|\bm{\psi}(z)\rangle},
\end{equation}
where $\bm{\psi^\dag}(z)=(\psi_x^\dag(z),\psi_y^\dag(z))$ is a new set
of fields different from $\bm{\psi}(z)$. Denoting
$\bm\Psi(z)=(\psi_x(z),\psi_x^\dag(z),\psi_y(z),\psi_y^\dag(z))$,
this expansion can be written in the following way:
\begin{equation}\label{A21}
\hat{\rho}(t)= \int P(\bm{\Psi};t)
\hat{\Lambda}(\bm{\Psi})d\mu(\bm{\Psi}),
\end{equation}
where the integration measure $d\mu(\bm{\Psi})$ means that the
integration is carried over all the possible fields $\psi_{s}$ and
$\psi_{s}^\dag$, $s=x,y$. Taking into account the definition
(\ref{A20}) one can show that,
\begin{subequations}\label{A22}
\begin{eqnarray}
\hat{\psi}_s(z)\hat{\Lambda}&=&\psi_s(z)\hat{\Lambda},\\
\hat{\psi}^\dag_s(z)\hat{\Lambda}&=&\left(\psi_s^\dag(z)+\frac{\delta}{\delta \psi_s(z)}\right)\hat{\Lambda},\\
\hat{\Lambda}\hat{\psi}^\dag_s(z)&=&\psi_s^\dag(z)\hat{\Lambda},\\
\hat{\Lambda}\hat{\psi}_s(z)&=&\left(\frac{\delta}{\delta
\psi^\dag_s(z)}+\psi_s(z)\right)\hat{\Lambda},
\end{eqnarray}
\end{subequations}
where $\delta/\delta \psi_s(z)$ and $\delta/\delta \psi_s^\dag(z)$
are functional derivatives. Eqs.~(\ref{A22}) generalize the
corresponding monomode identities of \cite{Drummond80}.

The $P$-function always exist and is positive for any density
operator. The $P$-function is useful for calculating normal
ordered moments:
\begin{equation}\label{A22b}
\langle (\hat{\psi}_s^\dag)^m(\hat{\psi}_{s'})^n\rangle=\int
(\psi_s^\dag)^m (\psi_{s'})^n P(\bm{\Psi};t) d\mu(\bm{\Psi}).
\end{equation}
In particular, Eq.~(\ref{A22b}) shows that $P$ is normalized to
unity: $1=\int P(\bm{\Psi};t) d\mu(\bm{\Psi})$. It can be
interpreted as a genuine probability density on the (infinite
dimensional) space sustained by the field $\psi_s(z)$ and
$\psi^\dag_s(z)$.

To obtain the time evolution of $P$, we insert the expansion
(\ref{A21}) into the Liouville equation (\ref{A19}) and find
\begin{widetext}
\begin{equation}\label{A23}
\int \frac{\partial P}{\partial t}
\hat{\Lambda}(\bm{\Psi})d\mu(\bm{\Psi})=\int d\mu(\bm{\Psi}) \int
dzP(\bm{\Psi};t)\left(C_k(\bm{\Psi})\frac{\delta}{\delta\Psi_k(z)}
+\frac{1}{2}D_{kl}(\bm{\Psi})\frac{\delta^2}{\delta\Psi_k(z)\delta\Psi_l(z)}\right)\hat{\Lambda}(\bm{\Psi}),
\end{equation}
\end{widetext}
where summation over $k$ and $l$ is implied. The $C_k$'s are the
components of a four-dimension drift vector $\bm{C}$, with
\begin{eqnarray}
C_1(\bm{\Psi})&=&-\omega'_x\frac{\partial \psi_x}{\partial
z}+i\frac{\omega''_x}{2}\frac{\partial^2\psi_x}{\partial
z^2}\nonumber\\
&\phantom{=}&+i\Theta(\psi^{\dag}_x\psi_x\psi_x+(1-B)\psi^\dag_y\psi_y\psi_x)\nonumber\\
&\phantom{=}&+i\Theta B\psi^\dag_x \psi_y^2
e^{-2i\Delta\beta_0z}\label{A24}
\end{eqnarray}
The $C_2$, $C_3$, and $C_4$ components have a similar form. $C_2$
is obtained from (\ref{A24}) by making the substitution (i)
$i\rightarrow -i$, $\psi_x\leftrightarrow\psi^{\dag}_x$, and
$\psi_y\leftrightarrow\psi^{\dag}_y$. $C_3$ is obtained by (ii)
exchanging $x$- and $y$-indexes in (\ref{A24}), and making the
subtitution $\Delta\beta_0\rightarrow-\Delta\beta_0$. To obtain
$C_4$, both substitutions (i) and (ii) must be performed. The
$D_{kl}$ are the elements of a symmetric diffusion matrix $\bm{D}$
that can be written in the form $\bm{D}=\bm{B}\bm{B}^{T}$, where
\begin{equation}
\bm{B}=\sqrt{i\Theta}
\begin{pmatrix}
\psi_x &0&\sqrt{B}\psi_y e^{-i\Delta\beta_0z}&0\\
0&i\psi^\dag_x &0&i\sqrt{B}\psi^\dag_y e^{i\Delta\beta_0z}\\
\psi_y &0&-\sqrt{B} \psi_x e^{i\Delta\beta_0z}&0\\
0&i\psi^\dag_y&0&-i\sqrt{B}\psi^{\dag}_xe^{-i\Delta\beta_0z}
\end{pmatrix}.
\end{equation}
Using Eqs.~(\ref{A22}), on can deduce from Eq.~(\ref{A23}) that
the $P(\bm{\Psi};t)$ verifies a functional Fokker-Planck equation
with a semi positive-definite diffusion matrix. We refer to
\cite{Kennedy88a} for a demonstration since Eq.~(\ref{A23}) has
the same structure as Eq.~(4.19) in \cite{Kennedy88a}. Because of
the semipositivity of the diffusion matrix, the positivity of $P$
is maintained during evolution.

The stochastic equations equivalent to the Fokker-Planck equation
for $P$ can be written in the following compact form:
\begin{equation}\label{A26}
\frac{\partial}{\partial t}
\Psi_k(z,t)=C_k(\bm{\Psi})+B_{kl}(\bm{\Psi}) \zeta_l(z,t),
\end{equation}
where $(k,l)\in \{1,2,3,4\}^2$, and $\zeta_l(z,t)$ are the
independent zero-mean Gaussian white noise random fields
introduced in Sec.~\ref{sec:3} and characterized by the second
order moments (\ref{moments}). The $\bm{C}$ vector gives the
deterministic evolution of the fields as predicted by the
classical theory of light. The way vacuum fluctuations modify the
classical evolution is determined by the structure of the $\bm{B}$
matrix. If one discards the stochastic terms, the fields
$\psi^\dag_s$ and $\psi_s$ appear to be just complex conjugated
of each other. However, when vacuum fluctuations are taken into
account, $\psi^\dag_s$ and $\psi_s$ must be treated as independent
fields that are only complex conjugate in mean.

As they stand, Eqs.~(\ref{A26}) seems to differ from
Eqs.~(\ref{SCNLS}). Actually, both writings are equivalent. To
highlight the equivalence we first notice that, according to the
instantaneous-power normalization of the $(A_s,A_s^\dag)$ fields,
one has the following relations:
\begin{subequations}\label{A27}
\begin{eqnarray}
A_s(z,t)&=&\sqrt{\hbar \omega_0 v_{gs0}}\psi_s(z,t),\\
A_s^\dag(z,t)&=&\sqrt{\hbar \omega_0 v_{gs0}}\psi^\dag_s(z,t).
\end{eqnarray}
\end{subequations}
Inserting (\ref{A27}) into (\ref{A26}), and noting that
$\omega'_s=v_{gs0}$ and $\Theta=\hbar\omega_0 v_{g0}^2\gamma$, we
find
\begin{widetext}
\begin{subequations}\label{A29}
\begin{eqnarray}
\frac{\partial A_x}{\partial z}+\frac{1}{v_{gx0}}\frac{\partial A_x}{\partial t}&=&
+i\frac{\omega''_x}{2v_{gx0}}\frac{\partial^2A_x}{\partial z^2}
+i\gamma \left[A_x^\dag A_x+(1-B)A_y^\dag A_y\right] A_x
+i\gamma B (A_y)^2A_x^\dag e^{-2i\Delta\beta_0z}\nonumber\\
&\phantom{=}&+\sqrt{i\gamma \hbar \omega_0}\left[\zeta_1 A_x+\sqrt{B} \zeta_3A_ye^{-i\Delta\beta_0 z}\right],\\
\frac{\partial A_x^\dag}{\partial z}+\frac{1}{v_{gx0}}\frac{\partial A_x^\dag}{\partial t}&=&
-i\frac{\omega''_x}{2v_{gx0}}\frac{\partial^2A_x^\dag}{\partial z^2}
-i\gamma \left[A_x^\dag A_x+(1-B)A_y^\dag A_y\right] A_x^\dag
-i\gamma B (A_y^\dag)^2A_x e^{+2i\Delta\beta_0z}\nonumber\\
&\phantom{=}&+\sqrt{-i\gamma \hbar \omega_0}\left[\zeta_2 A_x^\dag+\sqrt{B} \zeta_4A_y^\dag e^{+i\Delta\beta_0 z}\right],\\
\frac{\partial A_y}{\partial z}+\frac{1}{v_{gy0}}\frac{\partial A_y}{\partial t}&=&
+i\frac{\omega''_y}{2v_{gy0}}\frac{\partial^2A_y}{\partial z^2}
+i\gamma \left[A_y^\dag A_y+(1-B)A_x^\dag A_x\right] A_y
+i\gamma B (A_x)^2A_y^\dag e^{+2i\Delta\beta_0z}\nonumber\\
&\phantom{=}&+\sqrt{i\gamma \hbar \omega_0}\left[\zeta_1 A_y-\sqrt{B} \zeta_3A_xe^{+i\Delta\beta_0 z}\right],\\
\frac{\partial A_y^\dag}{\partial z}+\frac{1}{v_{gy0}}\frac{\partial A_y^\dag}{\partial t}&=&
-i\frac{\omega''_y}{2v_{gy0}}\frac{\partial^2A_y^\dag}{\partial z^2}
-i\gamma \left[A_y^\dag A_y+(1-B)A_x^\dag A_x\right] A_y^\dag
-i\gamma B (A_x^\dag)^2A_y e^{-2i\Delta\beta_0z}\nonumber\\
&\phantom{=}&+\sqrt{-i\gamma \hbar \omega_0}\left[\zeta_2 A_y^\dag-\sqrt{B} \zeta_4A_x^\dag e^{-i\Delta\beta_0
z}\right].
\end{eqnarray}
\end{subequations}
\end{widetext}
Eqs.~(\ref{A29}) differs from Eqs.~(\ref{SCNLS}) only in the first
term of the right member. For each axis, the group-velocity
dispersion parameter is $\beta_{2s}=-\omega''_s/v_{gs0}^3$. When
the typical pulse duration $T$ is such that $T/\beta_{2s}$ is much
bigger than the group-velocity $v_{gs0}$, which is the common
situation in fiber-optics, the following operator approximation
holds
\begin{equation}\label{A30}
\frac{\partial^2}{\partial z^2}\approx
\frac{1}{v_{gs0}^2}\frac{\partial^2}{\partial t^2}.
\end{equation}
Inserting (\ref{A30}) into Eqs.~(\ref{A29}), and noting that
usually $\beta_{2x}\approx \beta_{2y}\equiv \beta_2$ one obtains
the stochastic equations (\ref{SCNLS}).

\newpage 

\begin{thebibliography}{28}
\expandafter\ifx\csname
natexlab\endcsname\relax\def\natexlab#1{#1}\fi
\expandafter\ifx\csname bibnamefont\endcsname\relax
  \def\bibnamefont#1{#1}\fi
\expandafter\ifx\csname bibfnamefont\endcsname\relax
  \def\bibfnamefont#1{#1}\fi
\expandafter\ifx\csname citenamefont\endcsname\relax
  \def\citenamefont#1{#1}\fi
\expandafter\ifx\csname url\endcsname\relax
  \def\url#1{\texttt{#1}}\fi
\expandafter\ifx\csname
urlprefix\endcsname\relax\def\urlprefix{URL }\fi
\providecommand{\bibinfo}[2]{#2}
\providecommand{\eprint}[2][]{\url{#2}}

\bibitem[{\citenamefont{Levenson
  et~al.}(1985{\natexlab{a}})\citenamefont{Levenson, Shelby, Aspect, Reid, and
  Walls}}]{Levenson85a}
\bibinfo{author}{\bibfnamefont{M.~D.} \bibnamefont{Levenson}},
  \bibinfo{author}{\bibfnamefont{R.~M.} \bibnamefont{Shelby}},
  \bibinfo{author}{\bibfnamefont{A.}~\bibnamefont{Aspect}},
  \bibinfo{author}{\bibfnamefont{M.}~\bibnamefont{Reid}}, \bibnamefont{and}
  \bibinfo{author}{\bibfnamefont{D.~F.} \bibnamefont{Walls}},
  \bibinfo{journal}{Phys. Rev. A} \textbf{\bibinfo{volume}{32}},
  \bibinfo{pages}{1550} (\bibinfo{year}{1985}{\natexlab{a}}).

\bibitem[{\citenamefont{Levenson
  et~al.}(1985{\natexlab{b}})\citenamefont{Levenson, Shelby, and
  Perlmutter}}]{Levenson85b}
\bibinfo{author}{\bibfnamefont{M.~D.} \bibnamefont{Levenson}},
  \bibinfo{author}{\bibfnamefont{R.~M.} \bibnamefont{Shelby}},
  \bibnamefont{and} \bibinfo{author}{\bibfnamefont{S.~H.}
  \bibnamefont{Perlmutter}}, \bibinfo{journal}{Opt. Lett.}
  \textbf{\bibinfo{volume}{10}}, \bibinfo{pages}{514}
  (\bibinfo{year}{1985}{\natexlab{b}}).

\bibitem[{\citenamefont{Sizmann and Leuchs}(1999)}]{Sizmann99}
\bibinfo{author}{\bibfnamefont{A.}~\bibnamefont{Sizmann}} \bibnamefont{and}
  \bibinfo{author}{\bibfnamefont{G.}~\bibnamefont{Leuchs}}, in
  \emph{\bibinfo{booktitle}{Progress in Optics 39}}, edited by
  \bibinfo{editor}{\bibfnamefont{E.}~\bibnamefont{Wolf}}
  (\bibinfo{publisher}{North-Holland}, \bibinfo{address}{Amsterdam},
  \bibinfo{year}{1999}).

\bibitem[{\citenamefont{Fiorentino et~al.}(2002)\citenamefont{Fiorentino, Voss,
  Sharping, and Kumar}}]{Fiorentino02}
\bibinfo{author}{\bibfnamefont{M.}~\bibnamefont{Fiorentino}},
  \bibinfo{author}{\bibfnamefont{P.~L.} \bibnamefont{Voss}},
  \bibinfo{author}{\bibfnamefont{J.~E.} \bibnamefont{Sharping}},
  \bibnamefont{and} \bibinfo{author}{\bibfnamefont{P.}~\bibnamefont{Kumar}},
  \bibinfo{journal}{IEEE Photonics Technology Letters}
  \textbf{\bibinfo{volume}{14}}, \bibinfo{pages}{983} (\bibinfo{year}{2002}).

\bibitem[{\citenamefont{Agrawal}(1995)}]{Agrawal95}
\bibinfo{author}{\bibfnamefont{G.~P.} \bibnamefont{Agrawal}},
  \emph{\bibinfo{title}{Nonlinear Fiber Optics}} (\bibinfo{publisher}{Academic
  Press}, \bibinfo{address}{San Diego}, \bibinfo{year}{1995}).

\bibitem[{\citenamefont{Potasek and Yurke}(1987)}]{Potasek87}
\bibinfo{author}{\bibfnamefont{M.~J.} \bibnamefont{Potasek}} \bibnamefont{and}
  \bibinfo{author}{\bibfnamefont{B.}~\bibnamefont{Yurke}},
  \bibinfo{journal}{Phys. Rev. A} \textbf{\bibinfo{volume}{35}},
  \bibinfo{pages}{3974} (\bibinfo{year}{1987}).

\bibitem[{\citenamefont{Carter et~al.}(1987)\citenamefont{Carter, Drummond,
  Reid, and Shelby}}]{Carter87}
\bibinfo{author}{\bibfnamefont{S.~J.} \bibnamefont{Carter}},
  \bibinfo{author}{\bibfnamefont{P.~D.} \bibnamefont{Drummond}},
  \bibinfo{author}{\bibfnamefont{M.~D.} \bibnamefont{Reid}}, \bibnamefont{and}
  \bibinfo{author}{\bibfnamefont{R.~M.} \bibnamefont{Shelby}},
  \bibinfo{journal}{Phys. Rev. Lett.} \textbf{\bibinfo{volume}{58}},
  \bibinfo{pages}{1841} (\bibinfo{year}{1987}).

\bibitem[{\citenamefont{Drummond and Carter}(1987)}]{Drummond87}
\bibinfo{author}{\bibfnamefont{P.~D.} \bibnamefont{Drummond}} \bibnamefont{and}
  \bibinfo{author}{\bibfnamefont{S.~J.} \bibnamefont{Carter}},
  \bibinfo{journal}{J. Opt. Soc. Amer.~B} \textbf{\bibinfo{volume}{4}},
  \bibinfo{pages}{1565} (\bibinfo{year}{1987}).

\bibitem[{\citenamefont{Kennedy and Wright}(1988)}]{Kennedy88a}
\bibinfo{author}{\bibfnamefont{T.~A.~B.} \bibnamefont{Kennedy}}
  \bibnamefont{and} \bibinfo{author}{\bibfnamefont{E.~M.}
  \bibnamefont{Wright}}, \bibinfo{journal}{Phys. Rev. A}
  \textbf{\bibinfo{volume}{38}}, \bibinfo{pages}{212} (\bibinfo{year}{1988}).

\bibitem[{\citenamefont{Kennedy and Wabnitz}(1988)}]{Kennedy88b}
\bibinfo{author}{\bibfnamefont{T.~A.~B.} \bibnamefont{Kennedy}}
  \bibnamefont{and} \bibinfo{author}{\bibfnamefont{S.}~\bibnamefont{Wabnitz}},
  \bibinfo{journal}{Phys. Rev. A} \textbf{\bibinfo{volume}{38}},
  \bibinfo{pages}{563} (\bibinfo{year}{1988}).

\bibitem[{\citenamefont{Kennedy}(1991)}]{Kennedy91}
\bibinfo{author}{\bibfnamefont{T.~A.~B.} \bibnamefont{Kennedy}},
  \bibinfo{journal}{Phys. Rev. A} \textbf{\bibinfo{volume}{44}},
  \bibinfo{pages}{2113} (\bibinfo{year}{1991}).

\bibitem[{\citenamefont{Drummond and Corney}(2001)}]{Drummond01a}
\bibinfo{author}{\bibfnamefont{P.~D.} \bibnamefont{Drummond}} \bibnamefont{and}
  \bibinfo{author}{\bibfnamefont{J.~F.} \bibnamefont{Corney}},
  \bibinfo{journal}{J. Opt. Soc. Am. B} \textbf{\bibinfo{volume}{18}},
  \bibinfo{pages}{139} (\bibinfo{year}{2001}).

\bibitem[{\citenamefont{Amans et~al.}(2004)\citenamefont{Amans, Brainis,
  Emplit, and Massar}}]{Amans}
\bibinfo{author}{\bibfnamefont{D.}~\bibnamefont{Amans}},
  \bibinfo{author}{\bibfnamefont{E.}~\bibnamefont{Brainis}},
  \bibinfo{author}{\bibfnamefont{Ph.}~\bibnamefont{Emplit}}, \bibnamefont{and}
  \bibinfo{author}{\bibfnamefont{S.}~\bibnamefont{Massar}},
  \bibinfo{note}{in preparation}.

\bibitem[{\citenamefont{Stolen and Bjorkholm}(1982)}]{Stolen82}
\bibinfo{author}{\bibfnamefont{R.~H.} \bibnamefont{Stolen}} \bibnamefont{and}
  \bibinfo{author}{\bibfnamefont{J.~E.} \bibnamefont{Bjorkholm}},
  \bibinfo{journal}{IEEE J. Quantum Electron.}
  \textbf{\bibinfo{volume}{QE-18}}, \bibinfo{pages}{1062}
  (\bibinfo{year}{1982}).

\bibitem[{\citenamefont{Tai et~al.}(1986{\natexlab{a}})\citenamefont{Tai,
  Hasegawa, and Tomita}}]{Tai86a}
\bibinfo{author}{\bibfnamefont{K.}~\bibnamefont{Tai}},
  \bibinfo{author}{\bibfnamefont{A.}~\bibnamefont{Hasegawa}}, \bibnamefont{and}
  \bibinfo{author}{\bibfnamefont{A.}~\bibnamefont{Tomita}},
  \bibinfo{journal}{Phys. Rev. Lett.} \textbf{\bibinfo{volume}{56}},
  \bibinfo{pages}{135} (\bibinfo{year}{1986}{\natexlab{a}}).

\bibitem[{\citenamefont{Tai et~al.}(1986{\natexlab{b}})\citenamefont{Tai,
  Tomita, Jewell, and Hasegawa}}]{Tai86b}
\bibinfo{author}{\bibfnamefont{K.}~\bibnamefont{Tai}},
  \bibinfo{author}{\bibfnamefont{A.}~\bibnamefont{Tomita}},
  \bibinfo{author}{\bibfnamefont{J.~L.} \bibnamefont{Jewell}},
  \bibnamefont{and} \bibinfo{author}{\bibfnamefont{A.}~\bibnamefont{Hasegawa}},
  \bibinfo{journal}{Appl. Phys. Lett.} \textbf{\bibinfo{volume}{49}},
  \bibinfo{pages}{236} (\bibinfo{year}{1986}{\natexlab{b}}).

\bibitem[{\citenamefont{Lai and Haus}(1989)}]{LaiHaus.I.89}
\bibinfo{author}{\bibfnamefont{Y.}~\bibnamefont{Lai}} \bibnamefont{and}
  \bibinfo{author}{\bibfnamefont{H.~A.} \bibnamefont{Haus}},
  \bibinfo{journal}{Phys. Rev. A} \textbf{\bibinfo{volume}{40}},
  \bibinfo{pages}{844} (\bibinfo{year}{1989}).

\bibitem[{\citenamefont{Wright}(1991)}]{Wright91}
\bibinfo{author}{\bibfnamefont{E.~M.} \bibnamefont{Wright}},
  \bibinfo{journal}{Phys. Rev. A} \textbf{\bibinfo{volume}{43}},
  \bibinfo{pages}{3836} (\bibinfo{year}{1991}).

\bibitem[{\citenamefont{Haus}(2000)}]{Haus00}
\bibinfo{author}{\bibfnamefont{H.~A.} \bibnamefont{Haus}},
  \emph{\bibinfo{title}{Electromagnetic Noise and Quantum Optical
  Measurements}} (\bibinfo{publisher}{Springer-Verlag},
  \bibinfo{address}{Berlin}, \bibinfo{year}{2000}).

\bibitem[{\citenamefont{Korolkova and Leuchs}(2001)}]{Perina01}
\bibinfo{author}{\bibfnamefont{N.}~\bibnamefont{Korolkova}} \bibnamefont{and}
  \bibinfo{author}{\bibfnamefont{G.}~\bibnamefont{Leuchs}}, in
  \emph{\bibinfo{booktitle}{Coherence and Statistics of Photons and Atoms}},
  edited by \bibinfo{editor}{\bibfnamefont{J.}~\bibnamefont{Pe\v{r}ina}}
  (\bibinfo{publisher}{John Wiley \& Sons}, \bibinfo{address}{New York},
  \bibinfo{year}{2001}).

\bibitem[{\citenamefont{Blow et~al.}(1990)\citenamefont{Blow, Loudon, Phoenix,
  and Shepherd}}]{Blow90}
\bibinfo{author}{\bibfnamefont{K.~J.} \bibnamefont{Blow}},
  \bibinfo{author}{\bibfnamefont{R.}~\bibnamefont{Loudon}},
  \bibinfo{author}{\bibfnamefont{S.~J.~D.} \bibnamefont{Phoenix}},
  \bibnamefont{and} \bibinfo{author}{\bibfnamefont{T.~J.}
  \bibnamefont{Shepherd}}, \bibinfo{journal}{Phys. Rev. A}
  \textbf{\bibinfo{volume}{42}}, \bibinfo{pages}{4102} (\bibinfo{year}{1990}).

\bibitem[{\citenamefont{Drummond and Gardiner}(1980)}]{Drummond80}
\bibinfo{author}{\bibfnamefont{P.~D.} \bibnamefont{Drummond}} \bibnamefont{and}
  \bibinfo{author}{\bibfnamefont{C.~W.} \bibnamefont{Gardiner}},
  \bibinfo{journal}{J. Phys. A} \textbf{\bibinfo{volume}{13}},
  \bibinfo{pages}{2353} (\bibinfo{year}{1980}).

\bibitem[{\citenamefont{Memyuk}(1987)}]{Menyuk87}
\bibinfo{author}{\bibfnamefont{C.~R.} \bibnamefont{Memyuk}},
  \bibinfo{journal}{IEEE J. Quantum Electron.} \textbf{\bibinfo{volume}{23}},
  \bibinfo{pages}{174} (\bibinfo{year}{1987}).

\bibitem[{\citenamefont{Mallat}(1999)}]{Mallat99}
\bibinfo{author}{\bibfnamefont{S.}~\bibnamefont{Mallat}},
  \emph{\bibinfo{title}{A Wavelet Tour of Signal Processing}}
  (\bibinfo{publisher}{Academic Press}, \bibinfo{address}{San Diego},
  \bibinfo{year}{1999}).

\bibitem[{\citenamefont{Hillery and Mlodinow}(1984)}]{Hillery84}
\bibinfo{author}{\bibfnamefont{M.}~\bibnamefont{Hillery}} \bibnamefont{and}
  \bibinfo{author}{\bibfnamefont{L.~D.} \bibnamefont{Mlodinow}},
  \bibinfo{journal}{Phys. Rev. A} \textbf{\bibinfo{volume}{30}},
  \bibinfo{pages}{1860} (\bibinfo{year}{1984}).

\bibitem[{\citenamefont{Drummond}(1990)}]{Drummond90}
\bibinfo{author}{\bibfnamefont{P.~D.} \bibnamefont{Drummond}},
  \bibinfo{journal}{Phys. Rev. A} \textbf{\bibinfo{volume}{42}},
  \bibinfo{pages}{6845} (\bibinfo{year}{1990}).

\bibitem[{\citenamefont{Boyd}(1992)}]{Boyd92}
\bibinfo{author}{\bibfnamefont{R.~W.} \bibnamefont{Boyd}},
  \emph{\bibinfo{title}{Nonlinear Optics}} (\bibinfo{publisher}{Academic
  Press}, \bibinfo{address}{Boston}, \bibinfo{year}{1992}).

\bibitem[{\citenamefont{Svirko and Zheludev}(1998)}]{Svirko98}
\bibinfo{author}{\bibfnamefont{Y.~P.} \bibnamefont{Svirko}} \bibnamefont{and}
  \bibinfo{author}{\bibfnamefont{N.~I.} \bibnamefont{Zheludev}},
  \emph{\bibinfo{title}{Polarization of Light in Nonlinear Optics}}
  (\bibinfo{publisher}{John Wiley \& sons}, \bibinfo{address}{Chichester},
  \bibinfo{year}{1998}).

\end{thebibliography}

\end{document}